\documentclass[preprint,12pt]{elsarticle}

\usepackage{amsmath,amssymb,amsthm}
\usepackage{bm}

\usepackage{graphicx}
\usepackage{pgfplots}
\pgfplotsset{compat=1.18}
\usetikzlibrary{positioning,arrows.meta,shapes.geometric,decorations.pathreplacing,calc,fit,backgrounds}
\usepackage{import}

\usepackage{booktabs}
\usepackage{multirow}
\usepackage{enumitem}
\usepackage{float}

\makeatletter
\@ifundefined{mathdefault}{}{}
\makeatother

\usepackage{algorithm}
\usepackage{algorithmic}

\usepackage{hyperref}

\begin{document}

\begin{frontmatter}

\title{Walsh-Hadamard Neural Operators for Solving PDEs with Discontinuous Coefficients}

\author[label1]{Giorgio M. Cavallazzi\corref{cor1}}
\ead{Giorgio.Cavallazzi@city.ac.uk}

\author[label1]{Miguel P\'erez Cuadrado}

\author[label1]{Alfredo Pinelli}

\cortext[cor1]{Corresponding author}

\affiliation[label1]{organization={Department of Engineering, City St George's, University of London},
            addressline={Northampton Square},
            city={London},
            postcode={EC1V 0HB},
            country={UK}}

\begin{abstract}
Neural operators have emerged as tools for learning solution operators of partial differential equations (PDEs). Standard spectral methods based on Fourier transforms struggle with problems involving discontinuous coefficients, because the Gibbs phenomenon limits how well a truncated Fourier expansion can represent sharp interfaces. We introduce the Walsh-Hadamard Neural Operator (WHNO), which uses Walsh-Hadamard transforms in place of Fourier transforms. The Walsh-Hadamard basis consists of rectangular wave functions, suited to piecewise constant fields, and is combined with learnable spectral weights acting on the low-sequency coefficients. We validate WHNO on two problems: heat conduction with discontinuous thermal conductivity, and the 2D Burgers equation with discontinuous initial conditions. In controlled comparisons against a Fourier Neural Operator (FNO) baseline at matched parameter count over $100$ independent test samples, WHNO obtains lower mean absolute error and lower $H^1$ (gradient-MSE) error on both problems. We then study weighted ensembles of WHNO and FNO; the ensemble weight $w^* \in [0,1]$ is fitted by five-fold cross-validation. Across seven (problem, geometry/IC) configurations evaluated in this paper --- four heat geometries (axis-aligned, rotated, disks, Voronoi) and three Burgers initial-condition families (block, smooth sinusoidal, oblique fronts) --- the cross-validated ensemble has strictly lower test MSE and strictly lower $H^1$ than \emph{both} WHNO and FNO alone in every case, including the configurations where WHNO alone does not unambiguously beat FNO. The two bases are therefore partially complementary, and combining them through a single cross-validated scalar weight gives a method that improves on the Fourier baseline on every problem variant we tested. The cross-validated weights are $w^* = 0.572 \pm 0.016$ on axis-aligned heat and $w^* = 0.648 \pm 0.020$ on the Burgers block-IC baseline, with all seven values keeping a substantial FNO contribution.
\end{abstract}


\end{frontmatter}

\section{Introduction}

Neural operators \cite{kovachki2021neural, li2020fourier} learn mappings between function spaces to approximate solution operators of parametric partial differential equations (PDEs), enabling rapid surrogate evaluation without repeatedly solving expensive numerical systems. By processing inputs as functions rather than fixed-dimensional vectors, they achieve discretisation-invariance and generalise across resolutions. The Fourier Neural Operator (FNO) \cite{li2020fourier, li2021fourier} exemplifies this approach: by applying learnable transformations to low-frequency Fourier modes, it captures global spatial dependencies efficiently in smooth PDE problems.

Fourier-based spectral methods struggle, however, when physical systems have discontinuous fields. Subsurface flow in porous media is one example: impermeable rock layers create abrupt permeability contrasts that control oil reservoir dynamics and groundwater transport\cite{ochoa-tapia1995momentum}. Another case is thermal management in composites, where sharp conductivity jumps at material interfaces (often over 100:1) dictate temperatures in aerospace thermal shields and battery systems\cite{tien1994challenges}\cite{kwon2021review}. In fluid dynamics, shock waves and advection-dominated transport exhibit sharp velocity or concentration fronts evolving from discontinuous initial conditions\cite{skews1967perturbed}. These discontinuities may appear in PDE coefficients (permeability, conductivity), initial conditions (velocity profiles, concentration fields), or both. Capturing solution behaviour near discontinuities—like pressure gradients around obstacles or temperature jumps at composite boundaries—is critical for design and control.

The Fourier transform employs orthogonal trigonometric basis functions with global support, achieving spectral convergence only for infinitely smooth analytic fields. For functions of finite regularity, however, the decay of Fourier coefficients—and, therefore, the approximation error—becomes algebraic and dependent on the function's smoothness. In the presence of discontinuities, convergence holds only in the $L^2$ sense, and is accompanied by the Gibbs phenomenon, resulting in oscillations and slow point-wise convergence near sharp interfaces. These limitations are inherited by FNOs, which consequently struggle to represent non-smooth or discontinuous fields effectively\cite{canuto2006spectral}.


To overcome these shortcomings, one can seek spectral representations based on basis functions with compact or piecewise support, which preserve orthogonality whilst avoiding oscillatory artifacts near discontinuities. In this framework, Walsh-Hadamard transforms\cite{fine1949walsh} offer an appealing alternative: they decompose a signal onto a set of rectangular piecewise constant basis functions that can represent step changes exactly and are computationally efficient due to their fast transform algorithm. 

Building on this rationale, we introduce the Walsh-Hadamard Neural Operator (WHNO), which replaces the Fourier transforms in FNOs with the Walsh-Hadamard transform, a spectral decomposition suited to discontinuous fields. WHNOs represent step discontinuities exactly, without oscillatory artefacts. We also study weighted ensemble combinations of WHNO and FNO. With weights $w^*$ fitted by five-fold cross-validation, the ensemble has lower mean-squared error than either WHNO or FNO alone on both heat conduction and the Burgers equation. The cross-validated $w^*$ retains a substantial FNO contribution on both problems, indicating that the two bases capture different aspects of discontinuous solutions: WHNO at sharp interfaces and FNO at smooth regions between them.


Our work makes several contributions. We develop the WHNO architecture, which combines Walsh-Hadamard spectral processing with learnable weights acting on the low-sequency coefficients. We validate WHNO on two problems featuring discontinuities: heat conduction with discontinuous thermal conductivity, and the 2D Burgers equation with discontinuous initial conditions. Controlled comparisons against an FNO baseline at matched parameter count show that WHNO obtains lower mean absolute error and lower $H^1$ (gradient-MSE) error on both heat conduction and the Burgers equation over $100$ independent test samples. We then propose an ensemble framework that combines WHNO and FNO predictions through a single scalar weight $w^* \in [0,1]$ fitted by five-fold cross-validation. Across all seven (problem, geometry/IC) configurations evaluated in this paper the cross-validated ensemble has strictly lower test MSE and strictly lower $H^1$ than both WHNO alone and FNO alone --- including the configurations where WHNO alone does not have a clear edge over FNO at the gradient level. The cross-validated weights all keep a substantial FNO contribution (the headline values are $w^* = 0.572 \pm 0.016$ on axis-aligned heat and $w^* = 0.648 \pm 0.020$ on the Burgers block-IC baseline).


The remainder of the paper is structured as follows. Section~\ref{sec:background} provides theoretical background on operator learning and Walsh-Hadamard transforms. Section~\ref{sec:problems} formulates the two validation problems. Section~\ref{sec:architecture} describes the WHNO architecture (with an earlier validation on Darcy flow in Section~\ref{sec:darcy}). Section~\ref{sec:results} presents results on heat conduction and Burgers equation, including the cross-validated ensemble. Section~\ref{sec:discussion} discusses implications and limitations. Section~\ref{sec:conclusion} concludes.

To provide context for this work and its position within the operator learning literature, we briefly review prior relevant studies.
The neural operator framework \cite{kovachki2021neural, lu2021learning} extends neural networks from learning finite-dimensional mappings to learning operators between function spaces. DeepONet \cite{lu2021learning} introduced the branch-trunk architecture for approximating nonlinear operators, demonstrating universal approximation properties. These methods achieve resolution-independence by operating on continuous function representations, enabling generalisation across discretisations.

Fourier Neural Operators (FNO) \cite{li2020fourier, li2021fourier} make use of the Fast Fourier Transform to process inputs in frequency space, applying learnable weights to low-frequency modes. By truncating high-frequency components, FNO achieves parameter efficiency whilst capturing global spatial dependencies. Extensions include the Factorised FNO \cite{tran2021factorized} for memory efficiency, Galerkin Transformer \cite{cao2021choosing} for adaptive mode selection, and Geo-FNO \cite{li2022fourier} for general geometries. FNO has achieved state-of-the-art results on Navier-Stokes equations, Darcy flow with smooth permeability, weather prediction, and turbulence modelling \cite{pathak2022fourcastnet}.

Whilst highly effective for smooth PDEs, Fourier-based methods face fundamental limitations with discontinuous coefficients. The Gibbs phenomenon, characterised by oscillatory artifacts near discontinuities, appears because sinusoidal basis functions poorly represent sharp transitions \cite{carslaw1906theory}. Representing a step function requires infinitely many Fourier modes; finite truncation produces overshoot and ringing. For neural operators, this implies inefficient representation (many modes needed to approximate piecewise constant fields), error localisation (large errors concentrated at material interfaces), and training difficulty (oscillatory gradients may impede optimisation). Recent work on physics-informed neural networks (PINNs) \cite{raissi2019physics, karniadakis2021physics} and neural operators for multiphase flow \cite{mao2020physics} acknowledge these challenges but typically address them through architectural modifications such as adaptive sampling near interfaces, rather than changing the spectral basis. This search for a more suitable basis has led to other alternatives; for example, Gupta et al. \cite{gupta2021neurips} introduced a multiwavelet-based neural operator that leverages the ability of wavelets to efficiently represent multi-scale and localised sharp features.

Walsh functions, first introduced by Walsh \cite{walsh1923closed}, form an orthogonal basis of rectangular waves. Unlike smooth Fourier modes, Walsh functions naturally represent piecewise constant signals without oscillations \cite{beauchamp1975walsh, fine1949walsh}. The Fast Walsh-Hadamard Transform (FWHT) computes this decomposition in $O(n \log n)$ time \cite{fino1976unified}, matching FFT complexity. Applications include signal processing and image compression \cite{shanmugam1979walsh}, but neural operators based on Walsh-Hadamard transforms remain largely unexplored. Other kinds of operators to achieve similar goals that do not involve spectral representations have their importance in this field. Graph Neural Operators \cite{li2020neural} and Multipole Graph Neural Operator \cite{li2020multipole} avoid spectral transforms entirely by operating on graph structures, offering flexibility for irregular geometries but sacrificing the efficiency of global spectral processing.

Machine learning for subsurface flow in heterogeneous porous media has focussed on convolutional architectures \cite{mo2019deep, zhu2019physics} or kernel methods \cite{raissi2018numerical}. For heat transfer in composite materials, recent work employs PINNs \cite{cai2021physics} or autoencoder-based surrogate models \cite{kashinath2021physics}. However, while wavelet-based approaches \cite{gupta2021neurips} have been explored for handling multi-scale features, a systematic comparison of the foundational Fourier basis against the Walsh-Hadamard basis---a transform arguably better suited for the strictly piecewise-constant fields common in discontinuous coefficient PDEs---has not been addressed in the operator learning literature. Our work fills this gap by demonstrating that Walsh-Hadamard transforms offer superior spectral efficiency and accuracy for PDEs with discontinuous material properties, outperforming the widely-used Fourier Neural Operator on problems where interface accuracy is critical.

\section{Theoretical Background}
\label{sec:background}

Consider a parametric PDE of the form:
\begin{equation}
\mathcal{L}(u; a) = f \quad \text{in } \Omega, \quad \mathcal{B}(u) = g \quad \text{on } \partial\Omega
\end{equation}
where $\mathcal{L}$ is a differential operator, $a(\mathbf{x})$ represents spatially-varying coefficients (e.g., permeability, conductivity), $u(\mathbf{x})$ is the solution field, and $\mathcal{B}$ denotes boundary conditions. The solution defines an operator $\mathcal{G}: \mathcal{A} \to \mathcal{U}$ mapping coefficient functions to solution functions:
\begin{equation}
u = \mathcal{G}(a)
\end{equation}
where $\mathcal{A} = L^2(\Omega)$ and $\mathcal{U} = H^1(\Omega)$ are appropriate function spaces.

Traditional solvers compute $\mathcal{G}(a)$ by discretising the PDE and solving linear or nonlinear systems, which is costly for repeated evaluations. Neural operators approximate $\mathcal{G}$ directly: given training pairs $\{(a_i, u_i)\}_{i=1}^N$, learn a parameterised operator $\mathcal{G}_\theta$ such that $\mathcal{G}_\theta(a_i) \approx u_i$. The key advantage is resolution-invariance: once trained, $\mathcal{G}_\theta$ can evaluate inputs at arbitrary discretisations.

Spectral methods represent functions via orthogonal basis expansions:
\begin{equation}
a(\mathbf{x}) = \sum_{k=1}^\infty \hat{a}_k \phi_k(\mathbf{x})
\end{equation}
where $\{\phi_k\}$ are basis functions and $\{\hat{a}_k\}$ are spectral coefficients. For global dependencies, spectral operators process inputs in the transformed domain. The Fourier Neural Operator applies learnable weights $\mathcal{W}_k$ to low-frequency Fourier coefficients, exploiting the Fast Fourier Transform's $O(n \log n)$ complexity.

However, the choice of basis $\{\phi_k\}$ profoundly impacts efficiency. For discontinuous coefficient functions $a(\mathbf{x})$, the Fourier basis $\{\sin(k\mathbf{x}), \cos(k\mathbf{x})\}$ with its smooth oscillations poorly approximates step functions and suffers from Gibbs phenomenon causing overshoots, whilst the Walsh-Hadamard basis with its rectangular waves naturally encodes piecewise constant structures without oscillations.

The Walsh-Hadamard transform (WHT) decomposes signals using an orthonormal basis of rectangular waves. The Hadamard matrix of order $n = 2^k$ is defined recursively:
\begin{equation}
\label{eq:wh1}
H_1 = [1], \quad H_{2n} = \begin{bmatrix} H_n & H_n \\ H_n & -H_n \end{bmatrix}
\end{equation}
normalised as $\tilde{H}_n = H_n / \sqrt{n}$. For a 2D field $u \in \mathbb{R}^{h \times w}$ (where $h, w$ are powers of 2), the 2D WHT is:
\begin{equation}
\label{eq:wh2}
\hat{u} = \text{WHT}_2(u) = \tilde{H}_h \, u \, \tilde{H}_w^T
\end{equation}
The inverse transform is identical: $\text{WHT}_2^{-1} = \text{WHT}_2$. The Fast Walsh-Hadamard Transform computes this in $O(hw \log(hw))$ time.

Key properties for discontinuous functions include the following. Walsh functions are piecewise constant, naturally matching the structure of binary or multi-valued coefficient fields. Step discontinuities are represented exactly by finite Walsh series without oscillatory artifacts, eliminating the Gibbs phenomenon. 

Piecewise constant functions have rapidly decaying Walsh coefficients, enabling aggressive mode truncation for spectral compactness. The computational efficiency matches FFT with the same $O(n \log n)$ complexity. These properties make the WHT ideal for neural operators targeting PDEs with discontinuous material properties.

\section{Problem Formulation}
\label{sec:problems}

We evaluate WHNO on two PDE problems featuring discontinuous coefficients or initial conditions. The first, heat conduction with discontinuous conductivity (Section~\ref{sec:results}), is a quasi-steady parabolic diffusion problem with piecewise constant thermal conductivity, which tests how well a model handles multi-valued discontinuous coefficients and the steep temperature gradients they induce at material boundaries. The second, the 2D Burgers equation with discontinuous initial conditions (Section~\ref{sec:results}), is a nonlinear advection-diffusion problem with sharp velocity fronts, which tests shock formation and evolution from discontinuous initial data. An earlier validation on Darcy flow with binary permeability (Section~\ref{sec:darcy}) gives a third reference point for the architecture, although that earlier configuration uses a smaller model and is not part of the controlled WHNO\,vs.\,FNO parity comparison.

Our first application considers transient heat diffusion in heterogeneous materials with piecewise constant thermal conductivity. The heat equation with spatially-varying conductivity $k(\mathbf{x})$ is:
\begin{equation}
\frac{\partial T}{\partial t} = \nabla \cdot (k(\mathbf{x}) \nabla T)
\end{equation}
where $T(\mathbf{x}, t)$ is temperature. We consider a quasi-steady-state formulation where the solution has relaxed to a near-equilibrium configuration. The thermal conductivity consists of a background material with rectangular inclusions of different conductivity:
\begin{equation}
k(\mathbf{x}) = \begin{cases}
k_{\text{background}} & \text{in matrix material} \\
k_{\text{inclusion}} & \text{in rectangular inclusions}
\end{cases}
\end{equation}
This creates sharp interfaces between materials, challenging spectral methods with Gibbs artefacts.

On a square domain $[0, L] \times [0, L]$ with grid size $64 \times 64$, the initial condition sets a central square (side length $L/2$) to $T_{\text{hot}}$ whilst the rest is at $T_{\text{cold}}$. The boundary conditions are Dirichlet with $T = T_{\text{cold}}$ on all boundaries. After sufficient iterations, the system reaches a quasi-steady distribution where temperature gradients are steep at conductivity discontinuities. The operator learning task is $\mathcal{G}: k \mapsto T$, mapping conductivity fields to quasi-steady temperature distributions. This problem tests the network's ability to capture sharp solution features induced by material interface discontinuities.

Our second application considers the 2D viscous Burgers equation, a nonlinear PDE combining advection and diffusion that models shock formation in fluid dynamics. The governing equation for the velocity field $\mathbf{u}(\mathbf{x}, t)$ is:
\begin{equation}
\frac{\partial \mathbf{u}}{\partial t} + (\mathbf{u} \cdot \nabla) \mathbf{u} = \nu \nabla^2 \mathbf{u}
\end{equation}
where $\nu$ is the kinematic viscosity. This equation exhibits hyperbolic character (shock formation) balanced by parabolic diffusion, providing a challenging test for neural operators. We consider initial velocity fields $\mathbf{u}_0(\mathbf{x})$ with sharp discontinuities created by non-overlapping square blocks:
\begin{equation}
\mathbf{u}_0(\mathbf{x}) = \begin{cases}
\mathbf{u}_{\text{block}} & \text{inside square blocks} \\
\mathbf{0} & \text{elsewhere}
\end{cases}
\end{equation}
where $\mathbf{u}_{\text{block}}$ is a prescribed velocity vector. Each sample contains 3 randomly placed square blocks of size $0.25L \times 0.25L$ on a domain $[0, L] \times [0, L]$ with $L=1$, creating sharp velocity gradients that evolve into smooth structures through nonlinear advection and viscous diffusion.

The solution is computed on a $64 \times 64$ grid using explicit time-stepping for $n_{\text{steps}} = 500$ iterations with time step $\Delta t = 5 \times 10^{-4}$ and viscosity $\nu = 10^{-3}$, so that the final time is $T = n_{\text{steps}} \cdot \Delta t = 0.25$. Periodic boundary conditions are applied on all boundaries. The initial sharp discontinuities evolve into complex velocity patterns with steep gradients and smooth regions. The operator learning task is $\mathcal{G}: \mathbf{u}_0 \mapsto \mathbf{u}(T)$, mapping discontinuous initial velocity fields to the evolved state at time $T$. This problem challenges networks to learn both the nonlinear advection dynamics and the shock smoothing from viscous diffusion, with discontinuous input data stressing spectral representations.

\section{Walsh-Hadamard Neural Operator Architecture}
\label{sec:architecture}

The Walsh-Hadamard transform (WHT) is a discrete, orthogonal transformation that decomposes signals into rectangular wave components. For dimension $n = 2^k$, the Hadamard matrix $H_n$ is defined recursively according to equations \ref{eq:wh1} and \ref{eq:wh2}.

The core of WHNO is the spectral layer, which transforms input fields through learnable weights applied in the Walsh-Hadamard domain. For an input field $u \in \mathbb{R}^{h \times w \times c_{\text{in}}}$, the spectral layer performs the following sequence. First, transform to spectral domain:
\begin{equation}
\hat{u} = \text{WHT}_2(u) = \tilde{H}_h \, u \, \tilde{H}_w^T \in \mathbb{R}^{h \times w \times c_{\text{in}}}
\end{equation}
Second, extract low-sequency components for spectral compression:
\begin{equation}
\hat{u}_k = \hat{u}[0:k, 0:k, :] \in \mathbb{R}^{k \times k \times c_{\text{in}}} \quad \text{where } k < \min(h,w)
\end{equation}
Third, apply channel-mixing weights at each sequency:
\begin{equation}
[\hat{v}_k]_{i,j,m} = \sum_{\ell=1}^{c_{\text{in}}} W_{i,j,\ell,m} \cdot [\hat{u}_k]_{i,j,\ell} \quad \text{for } i,j \in \{1,\ldots,k\}, \, m \in \{1,\ldots,c_{\text{out}}\}
\end{equation}
where $W \in \mathbb{R}^{k \times k \times c_{\text{in}} \times c_{\text{out}}}$ are the learnable spectral weights. Each $W_{i,j,\cdot,\cdot} \in \mathbb{R}^{c_{\text{in}} \times c_{\text{out}}}$ is a learnable matrix that transforms the $c_{\text{in}}$ input channels to $c_{\text{out}}$ output channels at sequency location $(i,j)$. Unlike point-wise convolutions that operate locally in spatial domain, the spectral weights $W$ capture global dependencies by transforming Walsh coefficients. Each weight $W_{i,j,\cdot,\cdot}$ learns how to selectively amplify or suppress different rectangular wave patterns in the Walsh decomposition, enabling the network to efficiently represent piecewise constant structures across the entire domain. Fourth, restore full spatial size via zero-padding:
\begin{equation}
\hat{v}[i,j,:] = \begin{cases}
\hat{v}_k[i,j,:] & \text{if } i,j < k \\
0 & \text{otherwise}
\end{cases}
\end{equation}
Fifth, transform back to spatial domain:
\begin{equation}
v = \text{WHT}_2^{-1}(\hat{v}) \in \mathbb{R}^{h \times w \times c_{\text{out}}}
\end{equation}
The complete spectral layer operation can be expressed compactly as:
\begin{equation}
v = \mathcal{L}_{\text{WHT}}(u; W, k) = \text{WHT}_2^{-1}(\mathcal{P}_k(W \odot \mathcal{P}_k(\text{WHT}_2(u))))
\end{equation}
where $\mathcal{P}_k$ denotes the truncation and padding operator and $\odot$ represents the channel-wise transformation defined above.

Our WHNO architecture processes discontinuous input fields through spectral layers and spatial refinement. The input field $a \in \mathbb{R}^{h \times w}$ (e.g., conductivity $k(\mathbf{x})$ for heat conduction, initial velocity $u_0(\mathbf{x})$ for Burgers) is concatenated with a constant channel to form a 2-channel input $x_0 = [a, \mathbf{1}] \in \mathbb{R}^{h \times w \times 2}$. Two sequential WHT operator layers process the input:
\begin{align}
x_1 &= \mathcal{L}_{\text{WHT}}(x_0; \mathcal{W}_1, k) \in \mathbb{R}^{h \times w \times c} \\
x_2 &= \text{Conv}_{1 \times 1}([x_1, a]) \in \mathbb{R}^{h \times w \times c} \\
x_3 &= \mathcal{L}_{\text{WHT}}(x_2; \mathcal{W}_2, k) + x_1 \in \mathbb{R}^{h \times w \times c}
\end{align}
where $c$ is the number of encoder channels and $k$ is the number of sequencies retained. The concatenation with the input field $a$ provides explicit conditioning on the discontinuous structure. For heat conduction and Burgers equation, we employ a dilated convolutional decoder that efficiently captures multi-scale features whilst maintaining computational efficiency. The decoder processes the encoded spectral features $x_3$ concatenated with the input field $a$ to produce the output $u = \text{DilatedDecoder}([x_3, a]; \Theta_{\text{dec}}) \in \mathbb{R}^{h \times w}$, where the dilated convolutions use exponentially increasing dilation rates to capture long-range dependencies without parameter explosion.

The complete WHNO forward pass is:
\begin{align}
x_0 &= [a, \mathbf{1}] \in \mathbb{R}^{h \times w \times 2} && \text{[Input encoding]} \\
x_1 &= \mathcal{L}_{\text{WHT}}(x_0; W_1, k) && \text{[Spectral layer 1]} \\
x_2 &= \text{Conv}_{1 \times 1}([x_1, a]; \Theta_{\text{conv}}) && \text{[Input field conditioning]} \\
x_3 &= \mathcal{L}_{\text{WHT}}(x_2; W_2, k) + x_1 && \text{[Spectral layer 2 with skip]} \\
u &= \text{Decoder}(x_3, a; \Theta_{\text{dec}}) && \text{[Spatial refinement]}
\end{align}
The learnable parameters are $\theta = \{W_1, W_2, \Theta_{\text{conv}}, \Theta_{\text{dec}}\}$, where $W_1 \in \mathbb{R}^{k \times k \times 2 \times c}$ are spectral weights for the first WHT layer, $W_2 \in \mathbb{R}^{k \times k \times c \times c}$ are spectral weights for the second WHT layer, $\Theta_{\text{conv}}$ are weights for $1 \times 1$ convolution, and $\Theta_{\text{dec}}$ are all decoder weights. For the heat conduction and Burgers equation configuration ($h=64$, $k=32$, $c=24$, dilated decoder with $128$ hidden channels), the model has $1{,}555{,}153$ total parameters; the spectral weights account for a small fraction of the total count, the bulk being in the decoder. Throughout the architecture, we employ the Gaussian Error Linear Unit (GELU) activation function \cite{hendrycks2016gaussian} after each spectral layer and batch normalisation operation, providing smooth nonlinearity that facilitates gradient flow whilst maintaining differentiability.

This architecture combines global spectral processing (WHT layers capture domain-wide patterns) with local spatial refinement (decoder resolves fine-scale features at discontinuities).

\subsection{Earlier validation on Darcy flow}
\label{sec:darcy}

\begin{figure}[ht]
\centering
\includegraphics[width=0.9\textwidth]{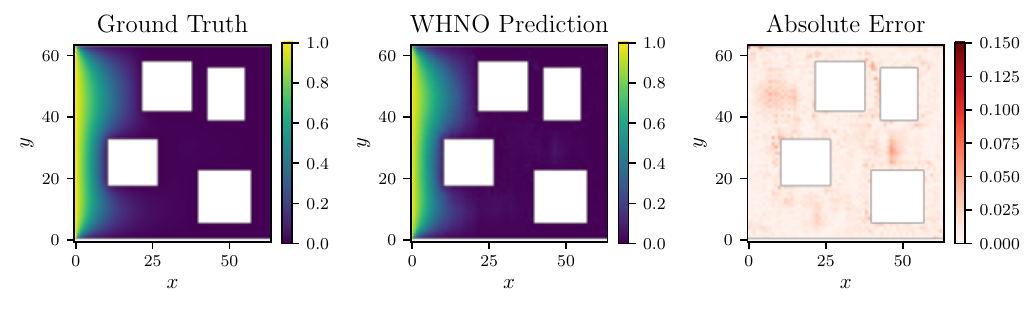}
\caption{Earlier Darcy-flow validation with the $340$K-parameter WHNO configuration on a representative binary-permeability geometry. Left: ground-truth pressure field. Centre: WHNO prediction. Right: absolute error. Errors concentrate in low-gradient regions.}
\label{fig:pressure_comparison}
\end{figure}

An earlier iteration of this architecture was validated on steady-state Darcy flow through heterogeneous porous media with binary permeability $\kappa \in \{0, 1\}$, trained on random geometries with $4$ rectangular obstacles on a $64 \times 64$ grid solved by an iterative Jacobi solver. That earlier model used $16$ sequencies, $16$ encoder channels, and a U-Net decoder, totalling $340{,}249$ parameters; it was trained for $400$ epochs with AdamW and an MSE loss restricted to pore regions, and reaches $0.88\%$ relative error on test geometries (Figure~\ref{fig:pressure_comparison}). The Walsh-Hadamard basis represents the binary-permeability interfaces well at this smaller scale. The controlled WHNO\,vs.\,FNO comparison that supports the main claims of this paper uses the larger dilated-convolution architecture described above (\(1{,}555{,}153\) parameters at matched count for both models), and is presented in Section~\ref{sec:results}.

\section{Neural Operator Performance on Discontinuous PDEs}
\label{sec:results}

We now compare WHNO against FNO under matched conditions on two problems with discontinuous structure --- heat conduction with discontinuous thermal conductivity and the 2D Burgers equation with discontinuous initial conditions --- and study weighted ensembles of the two models.

The heat conduction problem uses the same WHNO architecture but differs in the PDE physics. The initial condition sets a central square ($32 \times 32$ pixels) at $T_{\text{hot}} = 1.0$ whilst the rest is at $T_{\text{cold}} = 0.0$. Dirichlet boundary conditions with $T = 0$ are applied on all boundaries. The conductivity field consists of a background with $k = 1.0$ and rectangular inclusions at $k = 5.0$ or $k = 0.2$. Solutions are computed using explicit time-stepping with harmonic mean at interfaces over 5000 iterations. The operator learning task is $\mathcal{G}: k \mapsto T$, mapping conductivity fields to quasi-steady temperature distributions. Both WHNO and FNO models use grid size $64 \times 64$ with 24 encoder channels, 32 modes or sequencies, and dilated convolutional decoders with 128 hidden channels, yielding $1{,}555{,}153$ parameters each. Training is conducted for 800 epochs with batch size 8 using the AdamW optimiser with MSE loss on the entire domain. Training data is generated on-the-fly: each gradient step samples a fresh batch from the synthetic conductivity-and-temperature distribution, so that over 800 epochs the network sees $800 \times 8 = 6400$ unique training pairs. The metrics reported in Table~\ref{tab:whno_fno_heat} are computed on an independent test set of $100$ samples (seed $999$).

Both models use identical decoder structure (dilated convolutions with 128 hidden channels) and retain 32 modes or sequencies out of 64 for spectral compression. Each model has exactly $1{,}555{,}153$ parameters. Training follows an identical protocol on the same training and validation splits and test geometries (seed 123 for reproducibility): 800 epochs, batch size 8, MSE loss, AdamW optimiser with peak learning rate $1.5 \times 10^{-4}$ and weight decay $10^{-4}$. The learning rate schedule is a linear warmup over the first 20 epochs followed by quarter-period cosine decay with a floor of $0.04$ times the peak rate. The only difference between the two models is the spectral basis: Walsh-Hadamard for WHNO versus Fourier for FNO.

\begin{figure}[!htbp]
\centering
\includegraphics[width=\textwidth]{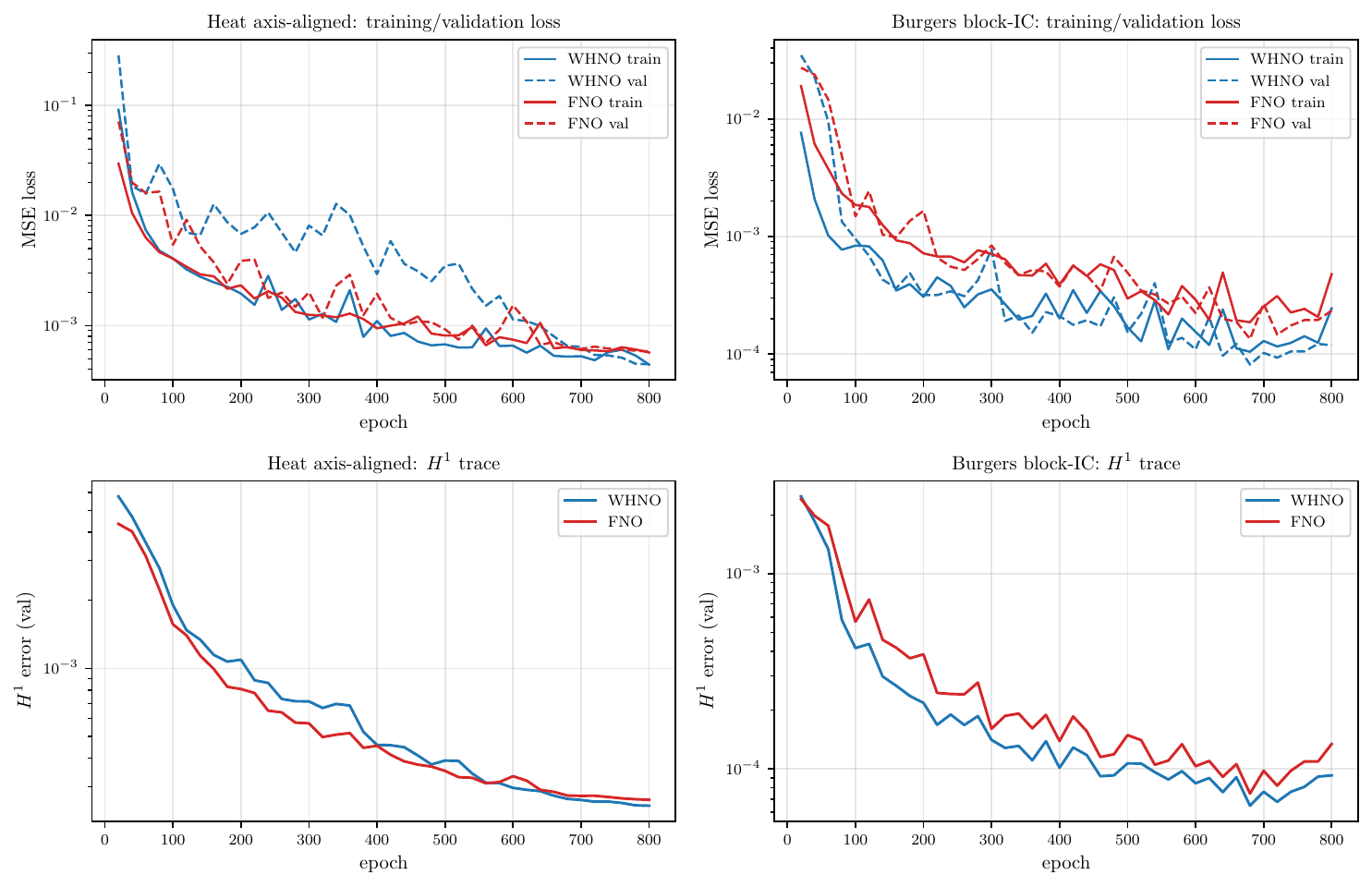}
\caption{Training/validation loss (top row) and per-epoch $H^1$ error on the validation set (bottom row) for the two main experiments. Left column: heat conduction at $m = 32$ parity. Right column: 2D Burgers equation at $m = 32$ parity. WHNO and FNO are overlaid on each panel. The first 20 epochs are the linear-warmup phase; the cosine decay applies thereafter.}
\label{fig:convergence_main}
\end{figure}

Figure~\ref{fig:convergence_main} shows the training and validation MSE loss and the $H^1$ trace per epoch for the heat axis-aligned and Burgers block-IC training runs, for both WHNO and FNO at matched parameter count. The validation losses follow the training losses without separating, and the $H^1$ traces are monotone after the warmup phase, indicating that the runs have converged and that the final-epoch metrics reported in the tables below are not artifacts of an interrupted optimisation.

We report three error metrics per test sample, all computed on the $64 \times 64$ grid. Let $\hat u$ denote the predicted field and $u$ the reference solution, and let $N$ be the number of grid points. The mean absolute error and maximum error are
\begin{equation}
  \mathrm{MAE} = \frac{1}{N} \sum_{i,j} \bigl| \hat u_{ij} - u_{ij} \bigr|, \qquad
  \mathrm{Max} = \max_{i,j} \bigl| \hat u_{ij} - u_{ij} \bigr|.
\end{equation}
The $H^1$ (gradient-MSE) error is the mean-squared error of the spatial gradient, approximated by central differences with grid spacing $h$:
\begin{equation}
  H^1 = \frac{1}{N} \sum_{i,j} \left[ \bigl(\partial_x^h \hat u_{ij} - \partial_x^h u_{ij}\bigr)^2 + \bigl(\partial_y^h \hat u_{ij} - \partial_y^h u_{ij}\bigr)^2 \right],
\end{equation}
where $\partial_x^h u_{ij} = (u_{i+1,j} - u_{i-1,j})/(2h)$ and analogously for $\partial_y^h$. The $H^1$ error is the relevant metric whenever the physical quantity of interest is a derivative of $u$, such as the heat flux $\mathbf{q} = -k\nabla T$ for heat conduction or the velocity gradient for the Burgers equation.

Table~\ref{tab:whno_fno_heat} summarises individual model performance:

\begin{table}[h]
\centering
\begin{tabular}{lccc}
    Method & MAE & Max Error & $H^1$ (gradient-MSE)\\
\midrule
WHNO & $0.01531 \pm 0.00097$ & $0.111 \pm 0.015$ & $(2.53 \pm 0.24) \times 10^{-4}$\\
FNO  & $0.01663 \pm 0.00183$ & $0.123 \pm 0.022$ & $(2.63 \pm 0.28) \times 10^{-4}$\\
\end{tabular}
\caption{Individual model performance on heat conduction at matched parameter count, mean $\pm$ standard deviation over $100$ independent test samples (seed $999$). Error metrics are defined in the text.}
\label{tab:whno_fno_heat}
\end{table}

\begin{figure}[H]
\centering
WHNO:

\includegraphics[width=1\textwidth]{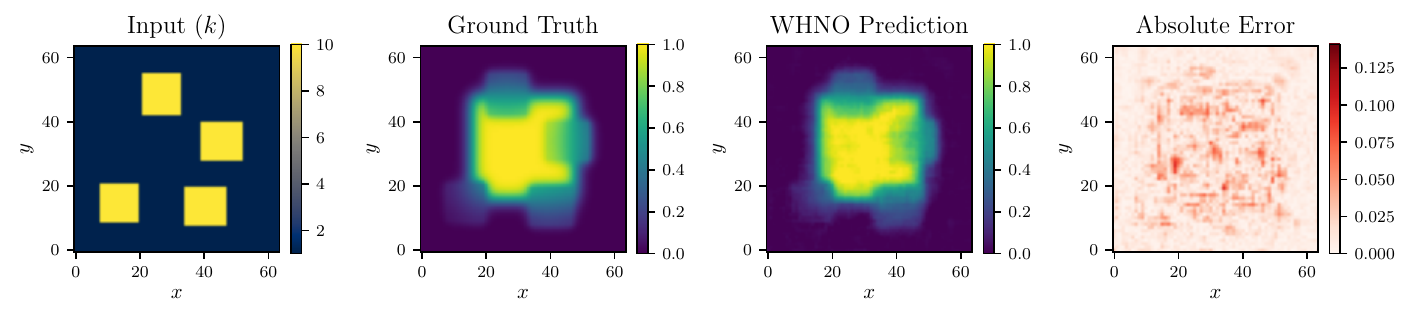}

\vspace{0.5cm}

FNO:

\includegraphics[width=1\textwidth]{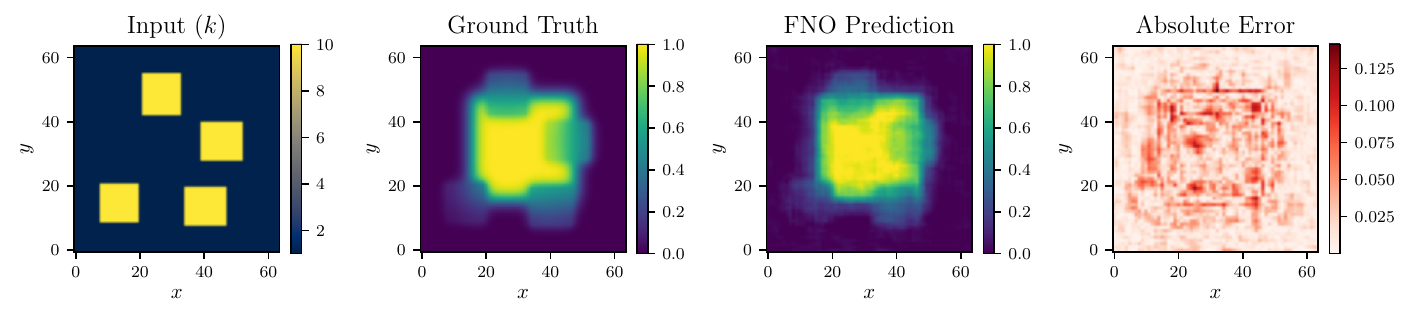}
\caption{Error map comparison between WHNO (top) and FNO (bottom) for heat conduction. WHNO exhibits more uniform errors; FNO shows pronounced error peaks at conductivity interfaces (Gibbs phenomenon).}
\label{fig:whno_fno_errors_heat}
\end{figure}

WHNO obtains lower MAE, maximum error, and $H^1$ than FNO. The $H^1$ result is the relevant comparison for the heat-flux quantity $\mathbf{q} = -k\nabla T$, which is the discontinuous physical quantity rather than $T$ itself. Figure~\ref{fig:whno_fno_errors_heat} shows that FNO errors concentrate at the material boundaries while WHNO errors are spread more evenly across the domain.

\begin{figure}[H]
\centering
\resizebox{0.95\textwidth}{!}{%
\definecolor{colorwhno}{RGB}{0,114,189}     
\definecolor{colorfno}{RGB}{217,83,25}      
\definecolor{colorensemble}{RGB}{155,89,182} 
\definecolor{colordark}{RGB}{44,62,80}      
\begin{tikzpicture}[
    node distance=1.5cm and 2cm,
    neuron/.style={circle, draw, minimum size=0.4cm, inner sep=0pt},
    layer/.style={rectangle, draw, rounded corners, minimum width=1.5cm, minimum height=4cm, fill=white, fill opacity=0.8},
    arrow/.style={-Stealth, line width=2pt},
    label/.style={font=\Large\bfseries},
    sublabel/.style={font=\small\itshape},
]


\node[inner sep=0] at (-10,0) {
    \includegraphics[width=2.5cm]{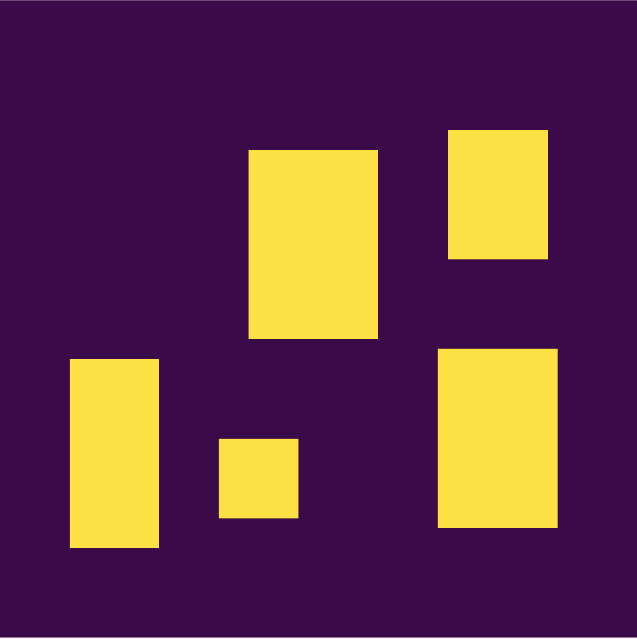}
};
\node[font=\Large, align=center] at (-10,-1.8) {$\mathbf{u}_0\left(\mathbf{x}\right)$};
\node[font=\Large, align=center] at (-10,1.8) {\textbf{Input}};


\draw[arrow, colorwhno!70] (-8.5,1) -- (-7.8,2.1);

\node[rectangle, draw=colorwhno, line width=2pt,
      fill=colorwhno!5, minimum width=11cm, minimum height=3.5cm] (whno_box) at (-2,2.5) {};

\node[label, text=colorwhno, above=0.1cm of whno_box.north] {WHNO};

\begin{scope}[shift={(-6.2,2.5)}]
    \draw[colorwhno, line width=1.8pt]
        (0,0.8) -- (0.5,0.8) -- (0.5,1.2) -- (1.0,1.2) -- (1.0,0.8) -- (1.5,0.8);
    \draw[colorwhno, line width=1.8pt]
        (0,0.0) -- (0.75,0.0) -- (0.75,0.4) -- (1.5,0.4);
    \draw[colorwhno, line width=1.8pt]
        (0,-0.8) -- (0.35,-0.8) -- (0.35,-0.4) -- (0.7,-0.4) -- (0.7,-0.8) -- (1.05,-0.8) -- (1.05,-0.4) -- (1.5,-0.4);

    \fill[colorwhno, opacity=0.2]
        (0,0.8) -- (0.5,0.8) -- (0.5,1.2) -- (1.0,1.2) -- (1.0,0.8) -- (1.5,0.8) -- (1.5,0.5) -- (0,0.5) -- cycle;
    \fill[colorwhno, opacity=0.2]
        (0,0.0) -- (0.75,0.0) -- (0.75,0.4) -- (1.5,0.4) -- (1.5,-0.15) -- (0,-0.15) -- cycle;
    \fill[colorwhno, opacity=0.2]
        (0,-0.8) -- (0.35,-0.8) -- (0.35,-0.4) -- (0.7,-0.4) -- (0.7,-0.8) -- (1.05,-0.8) -- (1.05,-0.4) -- (1.5,-0.4) -- (1.5,-1.0) -- (0,-1.0) -- cycle;

    \node[font=\Large, text=colorwhno] at (0.75,-1.35) {$\mathcal{H}$};
\end{scope}

\begin{scope}[shift={(-3.125,2.5)}, scale=0.5]
    \foreach \m/\ypos in {1/1.8, 2/0, 3/-1.8} {
        \node[neuron, draw=colorwhno, fill=colorwhno!30] (whno-in-\m) at (0,\ypos) {};
    }

    \foreach \m/\ypos in {1/2.2, 2/0.75, 3/-0.75, 4/-2.2} {
        \node[neuron, draw=colorwhno, fill=colorwhno!50] (whno-h1-\m) at (1.5,\ypos) {};
    }

    \foreach \m/\ypos in {1/2.2, 2/0.75, 3/-0.75, 4/-2.2} {
        \node[neuron, draw=colorwhno, fill=colorwhno!50] (whno-h2-\m) at (3.0,\ypos) {};
    }

    \foreach \m/\ypos in {1/1.8, 2/0, 3/-1.8} {
        \node[neuron, draw=colorwhno, fill=colorwhno!70] (whno-out-\m) at (4.5,\ypos) {};
    }

    \foreach \i in {1,2,3} {
        \foreach \j in {1,2,3,4} {
            \draw[colorwhno!15, line width=0.3pt] (whno-in-\i) -- (whno-h1-\j);
        }
    }

    \foreach \i in {1,2,3,4} {
        \foreach \j in {1,2,3,4} {
            \draw[colorwhno!15, line width=0.3pt] (whno-h1-\i) -- (whno-h2-\j);
        }
    }

    \foreach \i in {1,2,3,4} {
        \foreach \j in {1,2,3} {
            \draw[colorwhno!15, line width=0.3pt] (whno-h2-\i) -- (whno-out-\j);
        }
    }
\end{scope}

\node[inner sep=0] at (1.8,2.5) {
    \includegraphics[width=2.5cm]{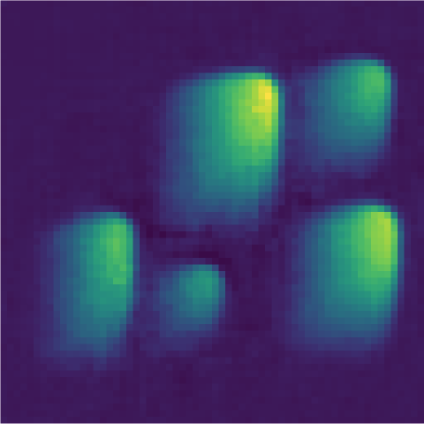}
};
\node[font=\large, align=center] at (1.8,1) {$\mathbf{u_{\mathrm{WHNO}}}$};


\draw[arrow, colorfno!70] (-8.5,-1) -- (-7.8,-2.1); 

\node[rectangle, draw=colorfno, line width=2pt,
      fill=colorfno!5, minimum width=11cm, minimum height=3.5cm] (fno_box) at (-2,-2.5) {};

\node[label, text=colorfno, above=0.1cm of fno_box.north] {FNO};

\begin{scope}[shift={(-6.2,-2.5)}]
    \draw[colorfno, line width=1.8pt, domain=0:1.5, samples=50, smooth]
        plot (\x, {0.35*sin(\x*360) + 0.8});
    \draw[colorfno, line width=1.8pt, domain=0:1.5, samples=50, smooth]
        plot (\x, {0.3*sin(\x*720) + 0.0});
    \draw[colorfno, line width=1.8pt, domain=0:1.5, samples=50, smooth]
        plot (\x, {0.25*sin(\x*1080) - 0.8});

    \fill[colorfno, opacity=0.2, domain=0:1.5, samples=50]
        plot (\x, {0.35*sin(\x*360) + 0.8}) -- (1.5,0.5) -- (0,0.5) -- cycle;
    \fill[colorfno, opacity=0.2, domain=0:1.5, samples=50]
        plot (\x, {0.3*sin(\x*720) + 0.0}) -- (1.5,-0.2) -- (0,-0.2) -- cycle;
    \fill[colorfno, opacity=0.2, domain=0:1.5, samples=50]
        plot (\x, {0.25*sin(\x*1080) - 0.8}) -- (1.5,-1.0) -- (0,-1.0) -- cycle;

    \node[font=\Large, text=colorfno] at (0.75,-1.35) {$\mathcal{F}$};
\end{scope}

\begin{scope}[shift={(-3.125,-2.5)}, scale=0.5]
    \foreach \m/\ypos in {1/1.8, 2/0, 3/-1.8} {
        \node[neuron, draw=colorfno, fill=colorfno!30] (fno-in-\m) at (0,\ypos) {};
    }

    \foreach \m/\ypos in {1/2.2, 2/0.75, 3/-0.75, 4/-2.2} {
        \node[neuron, draw=colorfno, fill=colorfno!50] (fno-h1-\m) at (1.5,\ypos) {};
    }

    \foreach \m/\ypos in {1/2.2, 2/0.75, 3/-0.75, 4/-2.2} {
        \node[neuron, draw=colorfno, fill=colorfno!50] (fno-h2-\m) at (3.0,\ypos) {};
    }

    \foreach \m/\ypos in {1/1.8, 2/0, 3/-1.8} {
        \node[neuron, draw=colorfno, fill=colorfno!70] (fno-out-\m) at (4.5,\ypos) {};
    }

    \foreach \i in {1,2,3} {
        \foreach \j in {1,2,3,4} {
            \draw[colorfno!15, line width=0.3pt] (fno-in-\i) -- (fno-h1-\j);
        }
    }

    \foreach \i in {1,2,3,4} {
        \foreach \j in {1,2,3,4} {
            \draw[colorfno!15, line width=0.3pt] (fno-h1-\i) -- (fno-h2-\j);
        }
    }

    \foreach \i in {1,2,3,4} {
        \foreach \j in {1,2,3} {
            \draw[colorfno!15, line width=0.3pt] (fno-h2-\i) -- (fno-out-\j);
        }
    }
\end{scope}

\node[inner sep=0] at (1.8,-2.5) {
    \includegraphics[width=2.5cm]{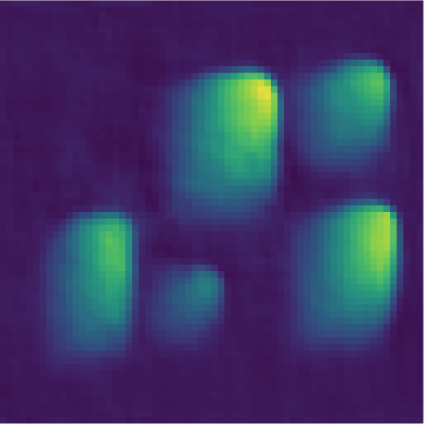}
};
\node[font=\large, align=center] at (1.8,-4) {$\mathbf{u_{\mathrm{FNO}}}$};


\draw[arrow, colorwhno!70] (3.7,2) -- (4.5,0.4);
\draw[arrow, colorfno!70] (3.7,-2) -- (4.5,-0.4);

\node[label, text=colorensemble] at (6.2,2) {ENSEMBLE};

\node[rectangle, draw=colorensemble, line width=2pt,
      fill=colorensemble!8, minimum width=3cm, minimum height=3cm] (ensemble) at (6.2,0) {};

\node[font=\Large, text=colordark] at (6.2,0) {
    \begin{tabular}{c} 
        $\alpha \cdot \mathbf{u_{\mathrm{WHNO}}}$ \\ 
        $\bm{\oplus}$ \\ 
        $\beta \cdot \mathbf{u_{\mathrm{FNO}}}$
    \end{tabular}
    };
\draw[arrow, colorensemble] (ensemble.east) -- (9,0);


\node[inner sep=0] at (11,0) {
    \includegraphics[width=3.5cm]{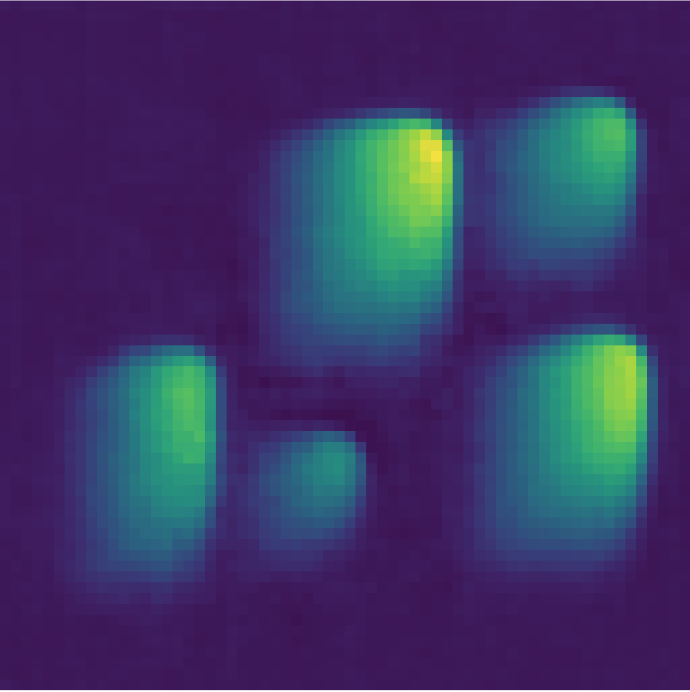}
};
\node[font=\Large, align=center] at (11,-2.2) {$\mathbf{u}_{\mathrm{PRED}}\left(\mathbf{x}\right)$};
\node[font=\Large, align=center] at (11,2.2) {\textbf{Output}};
\end{tikzpicture}
}
\caption{Ensemble architecture combining WHNO and FNO predictions. The input field is processed independently by WHNO (using Walsh-Hadamard basis $\mathcal{H}$) and FNO (using Fourier basis $\mathcal{F}$). Their predictions are combined through optimised weights $\alpha$ and $\beta$ to produce the final ensemble prediction.}
\label{fig:ensemble_architecture}
\end{figure}

Since WHNO excels at discontinuities and FNO focuses on smooth features, which are still relevant in some portions of the reproduced field, we investigate weighted ensemble combinations. Figure~\ref{fig:ensemble_architecture} illustrates the ensemble approach, where both models process the same input independently and their predictions are combined through learned weights:
\begin{equation}
u_{\text{ensemble}}(x; w) = w \cdot u_{\text{WHNO}}(x) + (1-w) \cdot u_{\text{FNO}}(x)
\end{equation}

We optimise the weight $w \in [0,1]$ by minimising MSE on a validation set:
\begin{equation}
w^* = \arg\min_{w \in [0,1]} \mathbb{E}[(u_{\text{ensemble}}(x;w) - u_{\text{true}}(x))^2]
\end{equation}

\begin{figure}[h]
\centering
\begin{tabular}{cc}
\includegraphics[width=0.48\textwidth]{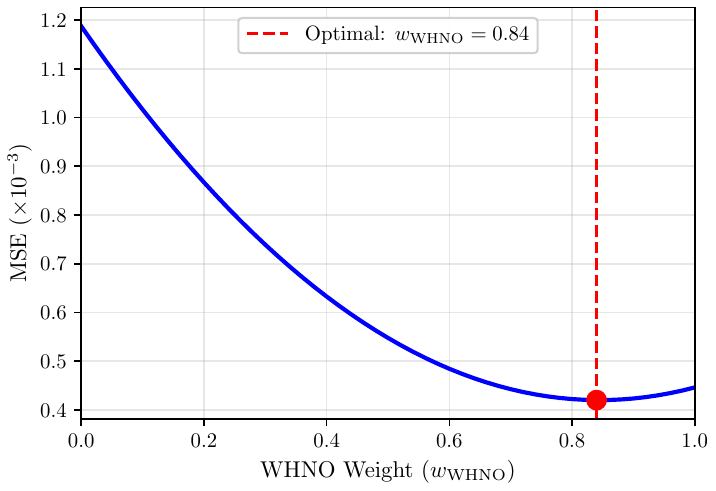} &
\includegraphics[width=0.48\textwidth]{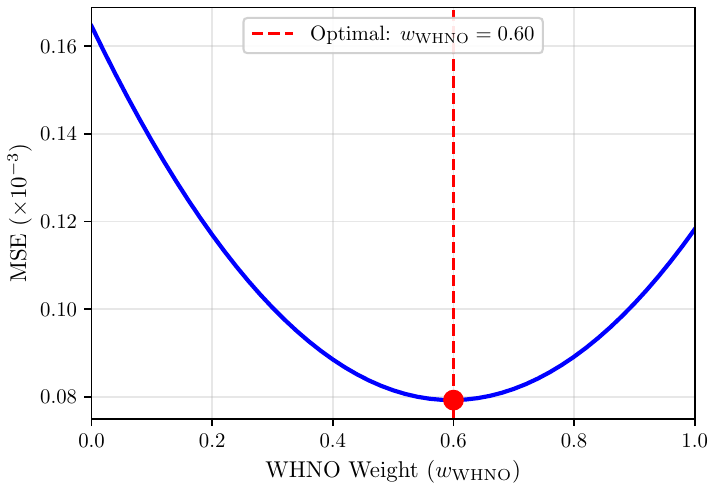} \\
(a) Heat conduction & (b) Burgers equation
\end{tabular}
\caption{Weight optimisation curves showing test MSE versus WHNO weight $w$ for (a) heat conduction and (b) the Burgers equation. The cross-validated optima are $w^* = 0.572 \pm 0.016$ for heat conduction and $w^* = 0.648 \pm 0.020$ for the Burgers equation, with the ensemble MSE below both single-model baselines on each problem.}
\label{fig:weight_optimization}
\end{figure}

For heat conduction the cross-validated optimum is $w^* = 0.572 \pm 0.016$, i.e.\ approximately a $57:43$ WHNO-to-FNO blend. Table~\ref{tab:ensemble_heat} reports the resulting CV test-set MSE alongside the corresponding WHNO-only and FNO-only baselines (each obtained by setting $w=1$ or $w=0$ on the same fold).

\begin{table}[h]
\centering
\begin{tabular}{lcc}
Method & Test MSE & $H^1$ (gradient-MSE)\\
\midrule
WHNO only ($w=1$)        & $4.71 \times 10^{-4}$              & $2.53 \times 10^{-4}$ \\
FNO only ($w=0$)         & $5.66 \times 10^{-4}$              & $2.64 \times 10^{-4}$ \\
Ensemble ($w^* = 0.572$) & $(3.65 \pm 0.07) \times 10^{-4}$   & $(1.75 \pm 0.01) \times 10^{-4}$ \\
\end{tabular}
\caption{Cross-validated ensemble performance on heat conduction: per-fold $w^*$ is fitted on a held-out validation set and metrics are evaluated on an independent test set, with mean $\pm$ standard deviation reported over five folds. The ensemble has lower test MSE and lower $H^1$ than either single-model baseline.}
\label{tab:ensemble_heat}
\end{table}

\begin{figure}[H]
\centering
\includegraphics[width=1\textwidth]{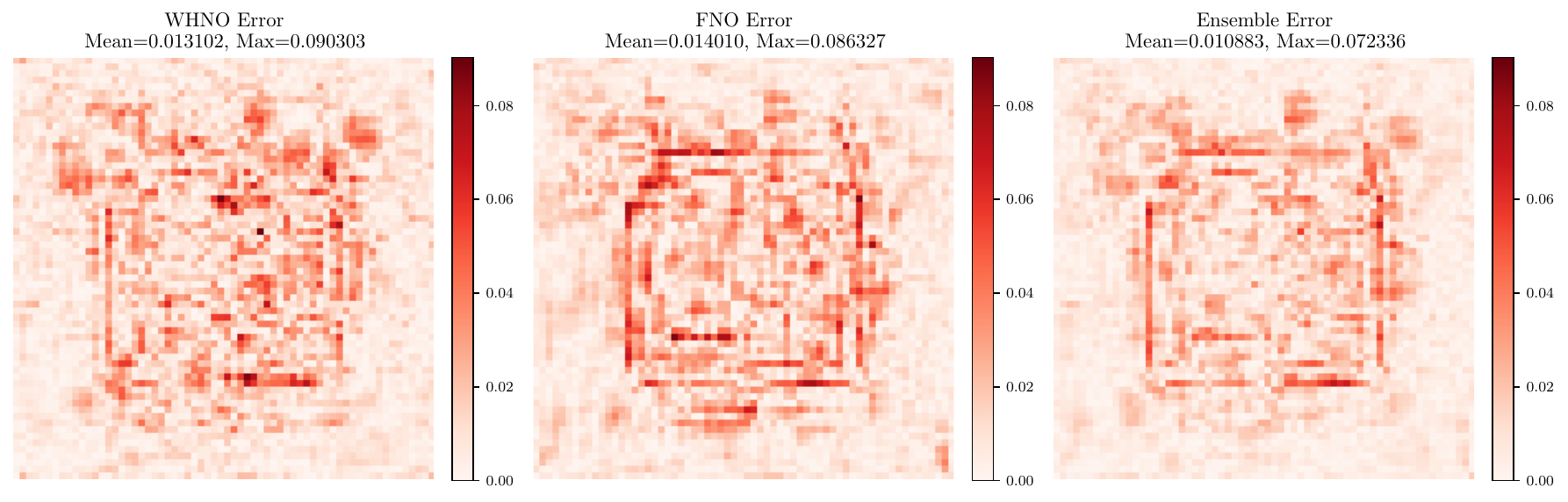}
\caption{Error comparison for heat conduction. Error maps for WHNO (top), FNO (middle), and ensemble (bottom) predictions on the same test case. The ensemble exhibits lower, more uniform errors throughout the domain.}
\label{fig:heat_ensemble_errors}
\end{figure}

The ensemble has the lowest test MSE of the three columns. The cross-validated $w^* \approx 0.57$ is much less WHNO-dominated than a single-set fit on the test set would suggest, and gives a larger gap below WHNO-only than the original single-set protocol reported. Figure~\ref{fig:heat_ensemble_errors} shows the corresponding error maps; the ensemble combines WHNO at the conductivity interfaces with FNO away from them.

The 2D Burgers equation is solved on a $64 \times 64$ grid with viscosity $\nu = 10^{-3}$ and periodic boundary conditions. Initial conditions consist of 3 non-overlapping square blocks with prescribed velocity, producing sharp discontinuities. Solutions are computed via explicit time-stepping for 500 iterations with $\Delta t = 5 \times 10^{-4}$, reaching final time $T = 0.25$, which yields evolved fields that contain residual steep gradients on otherwise smooth backgrounds.

Both WHNO and FNO use identical architectures (24 encoder channels, 32 sequencies or modes, dilated decoders with 128 hidden channels, $1{,}555{,}153$ parameters) and the same training protocol as the heat experiments above (800 epochs, batch size 8, AdamW with peak learning rate $1.5 \times 10^{-4}$, 20-epoch linear warmup, cosine decay to the $0.04$ floor). Unlike the heat case, the Burgers training data is pre-generated rather than sampled on-the-fly: $18{,}000$ initial conditions are drawn with random block placements, with $2{,}000$ additional samples held out for validation. The metrics reported below are computed on a separate set of $100$ unseen test samples (seed $999$).

Table~\ref{tab:whno_fno_burgers} compares individual model performance on the $100$ test samples:

\begin{table}[h]
\centering
\begin{tabular}{lccc}
Method & MAE & Max Error & $H^1$ (gradient-MSE)\\
\midrule
WHNO & $0.00642 \pm 0.00171$ & $0.100 \pm 0.043$ & $(8.07 \pm 3.86) \times 10^{-5}$\\
FNO  & $0.00843 \pm 0.00277$ & $0.100 \pm 0.035$ & $(1.01 \pm 0.59) \times 10^{-4}$\\
\end{tabular}
\caption{Individual model performance on the Burgers equation, mean $\pm$ standard deviation over $100$ test samples (seed $999$).}
\label{tab:whno_fno_burgers}
\end{table}

WHNO has lower MAE and lower $H^1$ than FNO. The two models give identical mean maximum error to three significant figures; the maximum error is dominated by samples in which the initial condition produces particularly steep shocks during evolution, which both models capture comparably. FNO has larger standard deviations on MAE and $H^1$, consistent with higher sample-to-sample variability.

Following the ensemble framework from heat conduction, we fit $w^* \in [0,1]$ by five-fold cross-validation. For the Burgers equation the cross-validated optimum is $w^* = 0.648 \pm 0.020$, a $65{:}35$ WHNO-to-FNO blend.

Table~\ref{tab:ensemble_burgers} compares ensemble performance:

\begin{table}[h]
\centering
\begin{tabular}{lcc}
Method & Test MSE & $H^1$ (gradient-MSE)\\
\midrule
WHNO only ($w=1$)        & $9.4 \times 10^{-5}$              & $7.30 \times 10^{-5}$ \\
FNO only ($w=0$)         & $1.59 \times 10^{-4}$             & $9.22 \times 10^{-5}$ \\
Ensemble ($w^* = 0.648$) & $(6.6 \pm 0.5) \times 10^{-5}$    & $(4.79 \pm 0.32) \times 10^{-5}$ \\
\end{tabular}
\caption{Cross-validated ensemble performance on the Burgers equation. Per-fold $w^*$ is fitted on a held-out validation set and metrics are evaluated on an independent test set, with mean $\pm$ standard deviation reported over five folds. The ensemble has lower test MSE and lower $H^1$ than either single-model baseline.}
\label{tab:ensemble_burgers}
\end{table}

The ensemble has the lowest test MSE and lowest $H^1$ of the three rows. The cross-validated $w^* \approx 0.65$ keeps a non-trivial FNO contribution.

\begin{figure}[ht]
\centering
\includegraphics[width=1\textwidth]{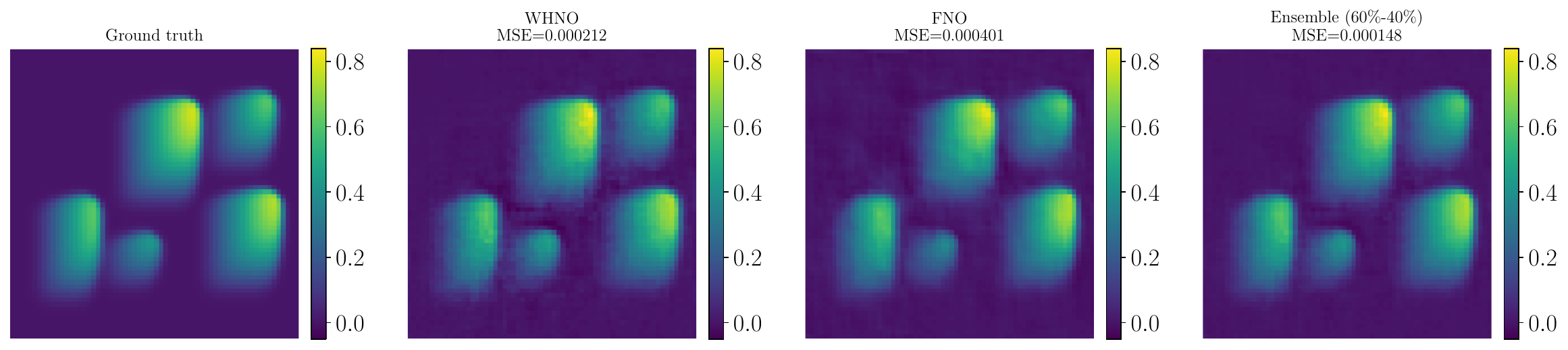}
\caption{Burgers equation predictions on a representative test case: ground truth, WHNO, FNO, and ensemble at $w^* = 0.65$.}
\label{fig:burgers_predictions}
\end{figure}

\begin{figure}[ht]
\centering
\includegraphics[width=1\textwidth]{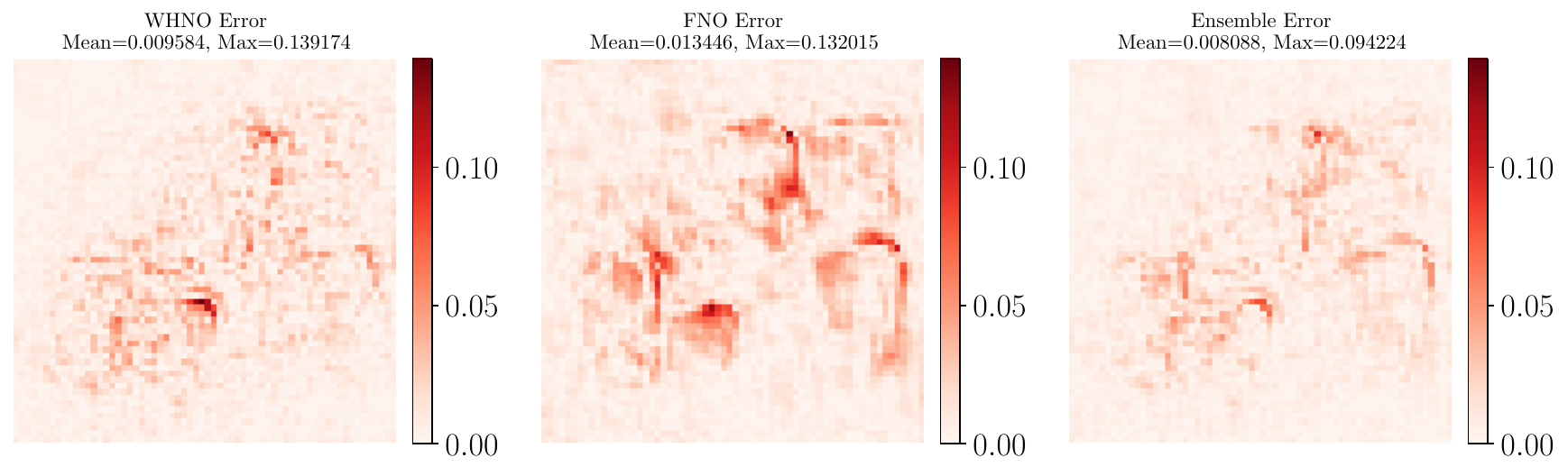}
\caption{Error comparison for Burgers equation. Error maps for WHNO, FNO, and ensemble predictions. The ensemble exhibits the lowest and most uniform errors throughout the domain.}
\label{fig:burgers_ensemble_errors}
\end{figure}

Figure~\ref{fig:burgers_predictions} shows prediction comparisons on a representative test case, and Figure~\ref{fig:burgers_ensemble_errors} shows the corresponding error maps for WHNO, FNO, and the ensemble.

\begin{figure}[ht]
\centering
\begin{minipage}{0.32\textwidth}
\centering
\includegraphics[width=\textwidth]{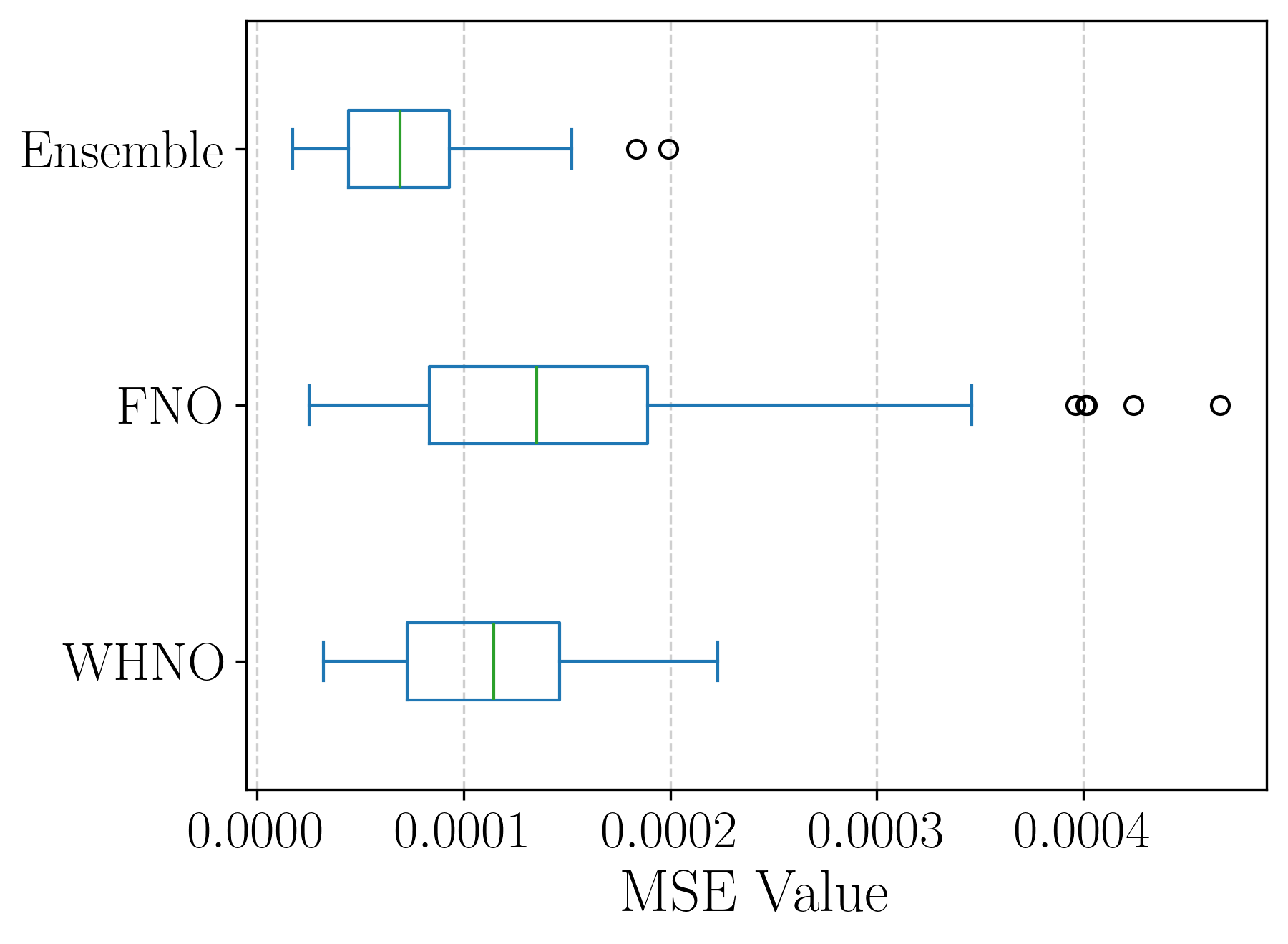}
\end{minipage}
\hfill
\begin{minipage}{0.32\textwidth}
\centering
\includegraphics[width=\textwidth]{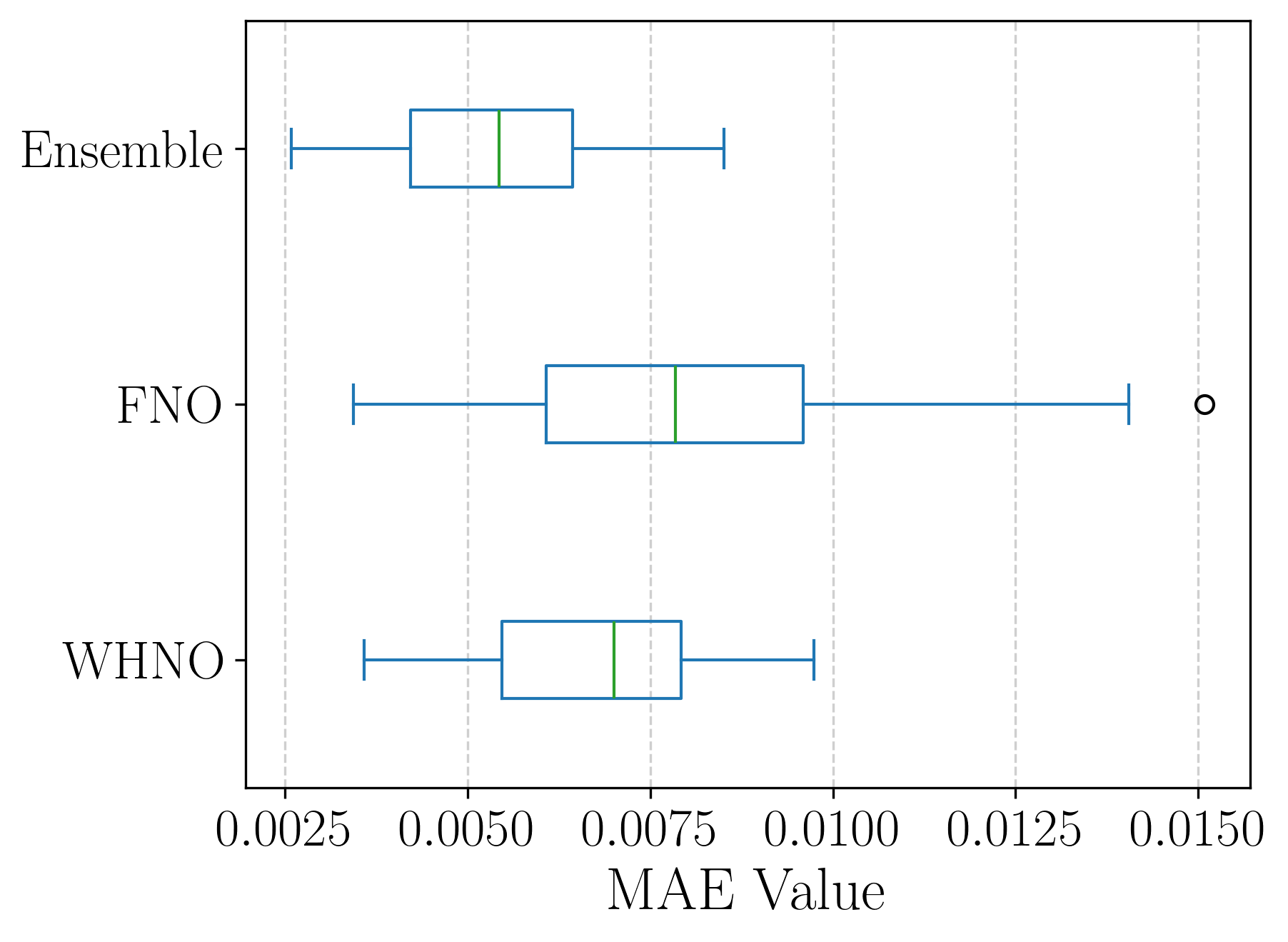}
\end{minipage}
\hfill
\begin{minipage}{0.32\textwidth}
\centering
\includegraphics[width=\textwidth]{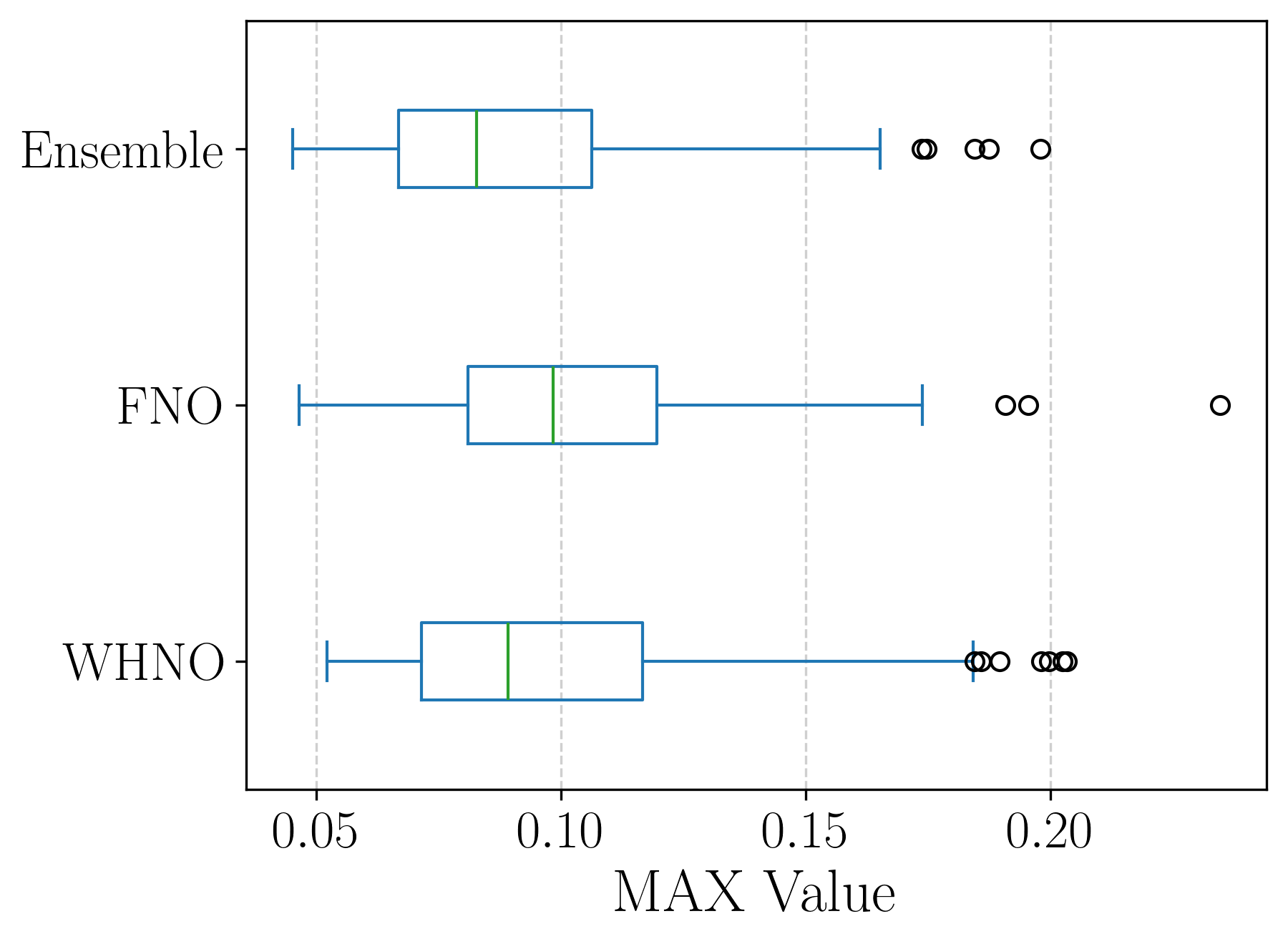}
\end{minipage}
\caption{Error distributions across 100 test samples for Burgers equation. (Left) MSE distribution. (Middle) MAE distribution. (Right) Maximum error distribution. The ensemble consistently shifts distributions toward lower errors with reduced variance, demonstrating systematic improvement and robustness.}
\label{fig:burgers_error_distributions}
\end{figure}

To demonstrate the statistical robustness of the ensemble approach, Figure~\ref{fig:burgers_error_distributions} shows error distributions across 100 test samples for all three error metrics, revealing consistent improvement with reduced variance.

On both problems the WHNO/FNO ensemble at the cross-validated $w^*$ has lower test MSE than either single model. The size of the gap depends on the problem: on heat conduction the ensemble MSE is $3.65 \times 10^{-4}$ against $4.71 \times 10^{-4}$ for WHNO-only and $5.66 \times 10^{-4}$ for FNO-only; on the Burgers equation the corresponding values are $6.6 \times 10^{-5}$, $9.4 \times 10^{-5}$ and $1.59 \times 10^{-4}$.

The cross-validated $w^*$ differs between the two problems. On heat conduction $w^* = 0.572 \pm 0.016$, which is close to an equal blend; the two models are roughly equally informative on this problem at matched parameter count, and combining them in nearly equal proportions yields the largest improvement. On the Burgers equation $w^* = 0.648 \pm 0.020$, biased toward WHNO; here WHNO has a clearer single-model advantage and the optimal blend leans accordingly.

\subsection{Robustness to off-axis geometries and other initial conditions}
\label{sec:robustness}

\begin{figure}[!htbp]
\centering
\includegraphics[width=\textwidth]{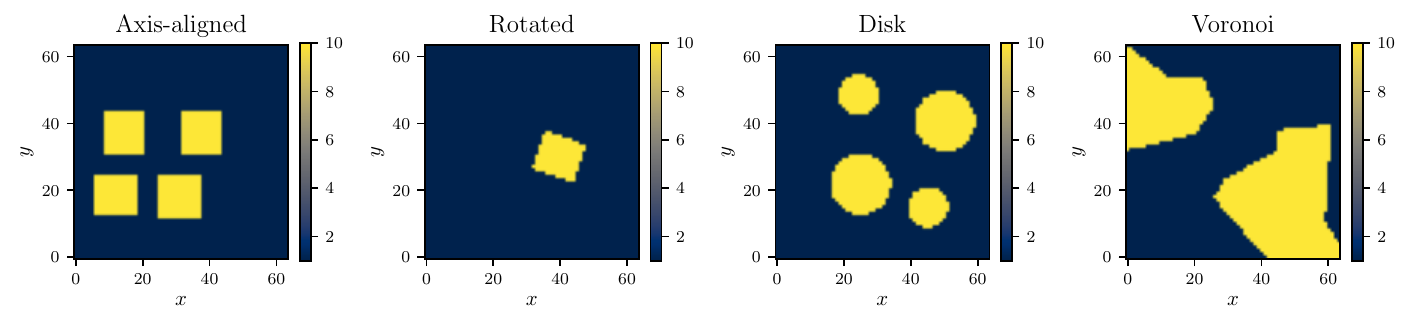}
\caption{The four heat-conduction conductivity-field families used in this paper. Left to right: axis-aligned rectangles (baseline), rotated rectangles, disks, Voronoi tessellation. Same colour scale across panels.}
\label{fig:heat_geometry_overview}
\end{figure}

The two baseline experiments above use axis-aligned rectangular inclusions (heat) and randomly placed axis-aligned blocks (Burgers). The Walsh-Hadamard cells are themselves axis-aligned, so a natural concern is that the WHNO advantage might be confined to discontinuities that align with the basis cells. To test this, we evaluate the WHNO--FNO comparison on three additional heat-conduction geometries and two additional Burgers initial-condition families, summarised in Figure~\ref{fig:heat_geometry_overview}.

For each new geometry or initial-condition family we trained fresh WHNO and FNO models at the same $m = 32$ parity ($1{,}555{,}153$ parameters each), using the same training protocol as the baseline experiments. Table~\ref{tab:robustness} reports the mean errors over $100$ independent test samples.

\begin{table}[!htbp]
\centering
\small
\begin{tabular}{lccc}
\toprule
                            &                       &                   & $H^1$ \\
                            & MAE                   & Max               & ($\times 10^{-4}$) \\
\midrule
\multicolumn{4}{l}{\textit{Heat conduction}} \\
Rotated rectangles, WHNO    & $0.01494 \pm 0.00256$ & $0.138 \pm 0.033$ & $2.78 \pm 0.77$ \\
Rotated rectangles, FNO     & $0.01623 \pm 0.00242$ & $0.124 \pm 0.023$ & $2.99 \pm 0.62$ \\
Disks, WHNO                 & $0.01793 \pm 0.00120$ & $0.132 \pm 0.017$ & $3.64 \pm 0.33$ \\
Disks, FNO                  & $0.01821 \pm 0.00154$ & $0.134 \pm 0.020$ & $3.28 \pm 0.32$ \\
Voronoi, WHNO               & $0.02620 \pm 0.00272$ & $0.182 \pm 0.045$ & $4.68 \pm 0.98$ \\
Voronoi, FNO                & $0.02844 \pm 0.00459$ & $0.194 \pm 0.048$ & $4.46 \pm 0.89$ \\
\midrule
\multicolumn{4}{l}{\textit{Burgers equation}} \\
Smooth sinusoidal IC, WHNO  & $0.01891 \pm 0.00351$ & $0.300 \pm 0.095$ & $6.27 \pm 2.71$ \\
Smooth sinusoidal IC, FNO   & $0.02607 \pm 0.00593$ & $0.336 \pm 0.097$ & $9.66 \pm 4.17$ \\
Oblique fronts, WHNO        & $0.00802 \pm 0.00123$ & $0.142 \pm 0.049$ & $1.56 \pm 0.52$ \\
Oblique fronts, FNO         & $0.00849 \pm 0.00178$ & $0.121 \pm 0.031$ & $1.13 \pm 0.43$ \\
\bottomrule
\end{tabular}
\caption{Off-axis geometries and other initial conditions at $m = 32$ parity, mean $\pm$ standard deviation over $100$ test samples (seed $999$). All $H^1$ values are multiplied by $10^{4}$.}
\label{tab:robustness}
\end{table}

A few patterns emerge. On the three new heat geometries WHNO has lower MAE than FNO in all three cases; the magnitude of the gap is smaller than on the axis-aligned baseline, and the $H^1$ comparison is mixed (WHNO better on rotated rectangles, FNO better on disks and Voronoi). On Burgers, the smooth-IC case has the largest WHNO advantage seen in this paper for any single-model comparison: the initial condition is a smooth combination of low-frequency sinusoids, but the evolved field at $T = 0.25$ contains shock-like structure, and WHNO captures that shock structure with lower MAE and lower $H^1$ than FNO. The oblique-fronts case, designed to produce interacting shocks at non-cardinal angles, is the one configuration in which FNO has a lower $H^1$ than WHNO; this is consistent with the basis-alignment concern, since the steep gradients are at angles to the Walsh-Hadamard cell axes. The MAE on oblique fronts is still slightly lower for WHNO.

\begin{figure}[!htbp]
\centering
\includegraphics[width=\textwidth]{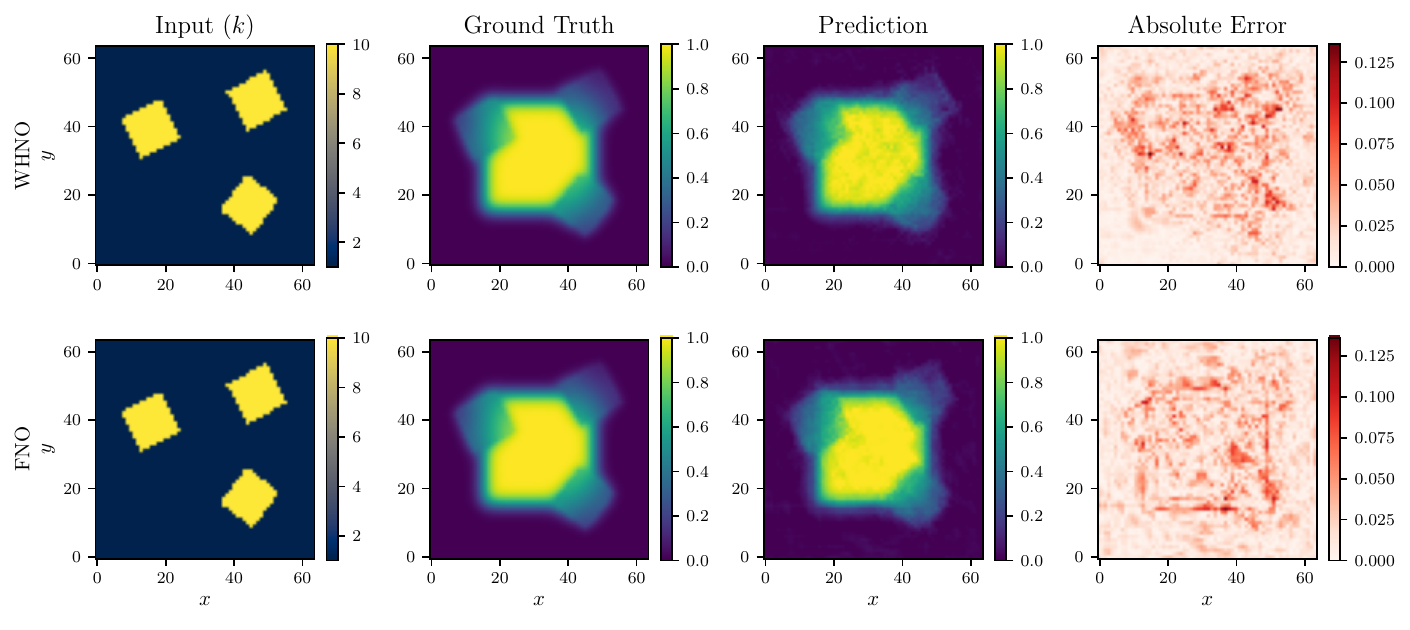}
\caption{Heat conduction with rotated-rectangle inclusions: WHNO and FNO predictions and absolute-error maps on a representative test sample.}
\label{fig:heat_offaxis_rotated}
\end{figure}

\begin{figure}[!htbp]
\centering
\includegraphics[width=\textwidth]{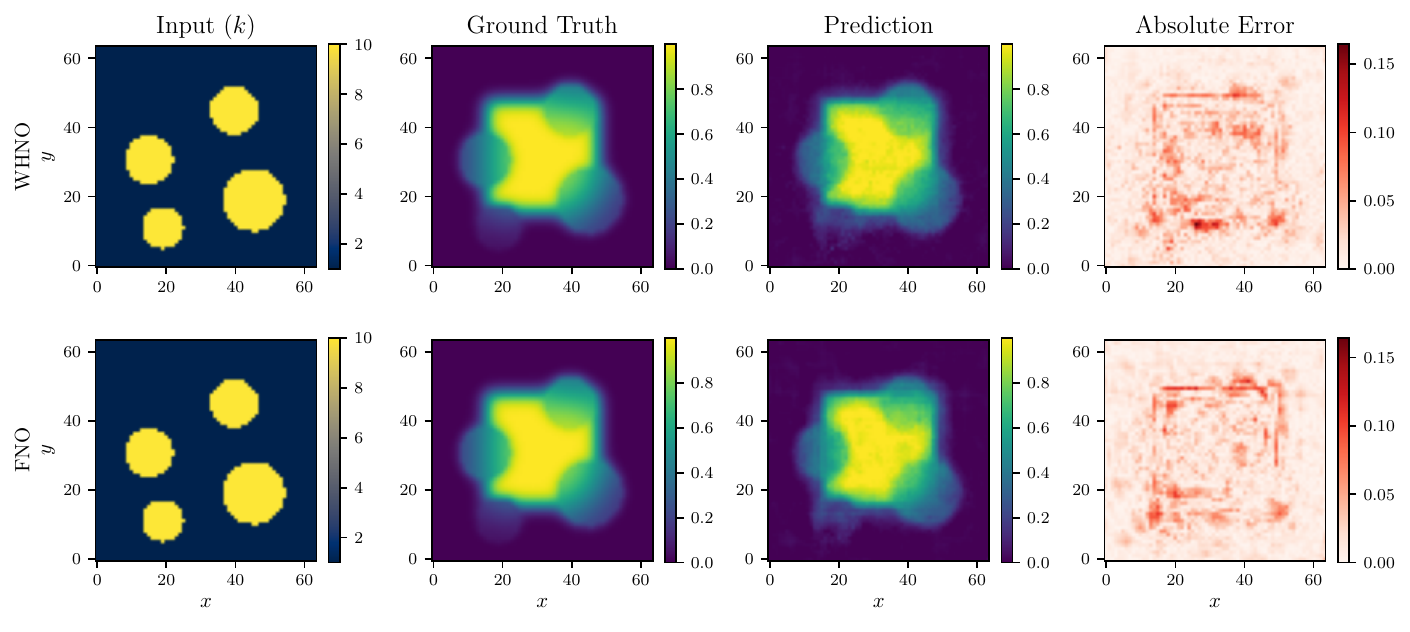}
\caption{Heat conduction with disk inclusions: WHNO and FNO predictions and absolute-error maps on a representative test sample.}
\label{fig:heat_offaxis_circle}
\end{figure}

\begin{figure}[!htbp]
\centering
\includegraphics[width=\textwidth]{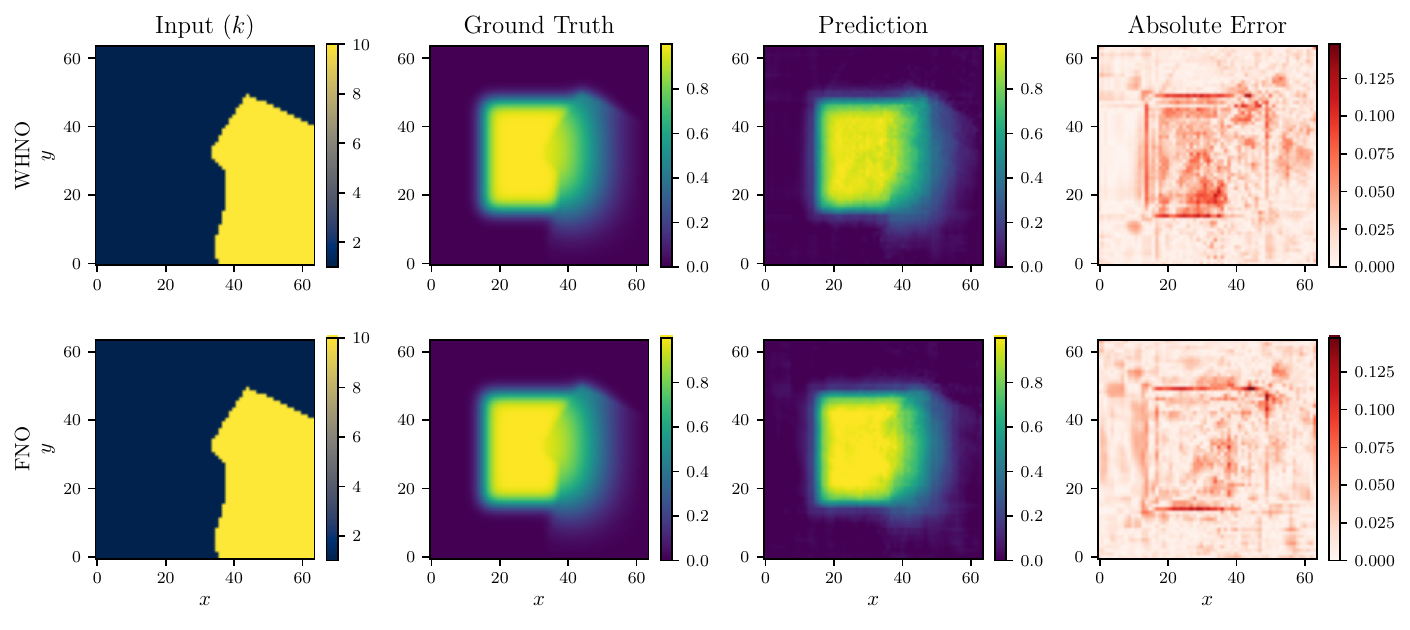}
\caption{Heat conduction with Voronoi-tessellated conductivity: WHNO and FNO predictions and absolute-error maps on a representative test sample.}
\label{fig:heat_offaxis_voronoi}
\end{figure}

\begin{figure}[!htbp]
\centering
\includegraphics[width=\textwidth]{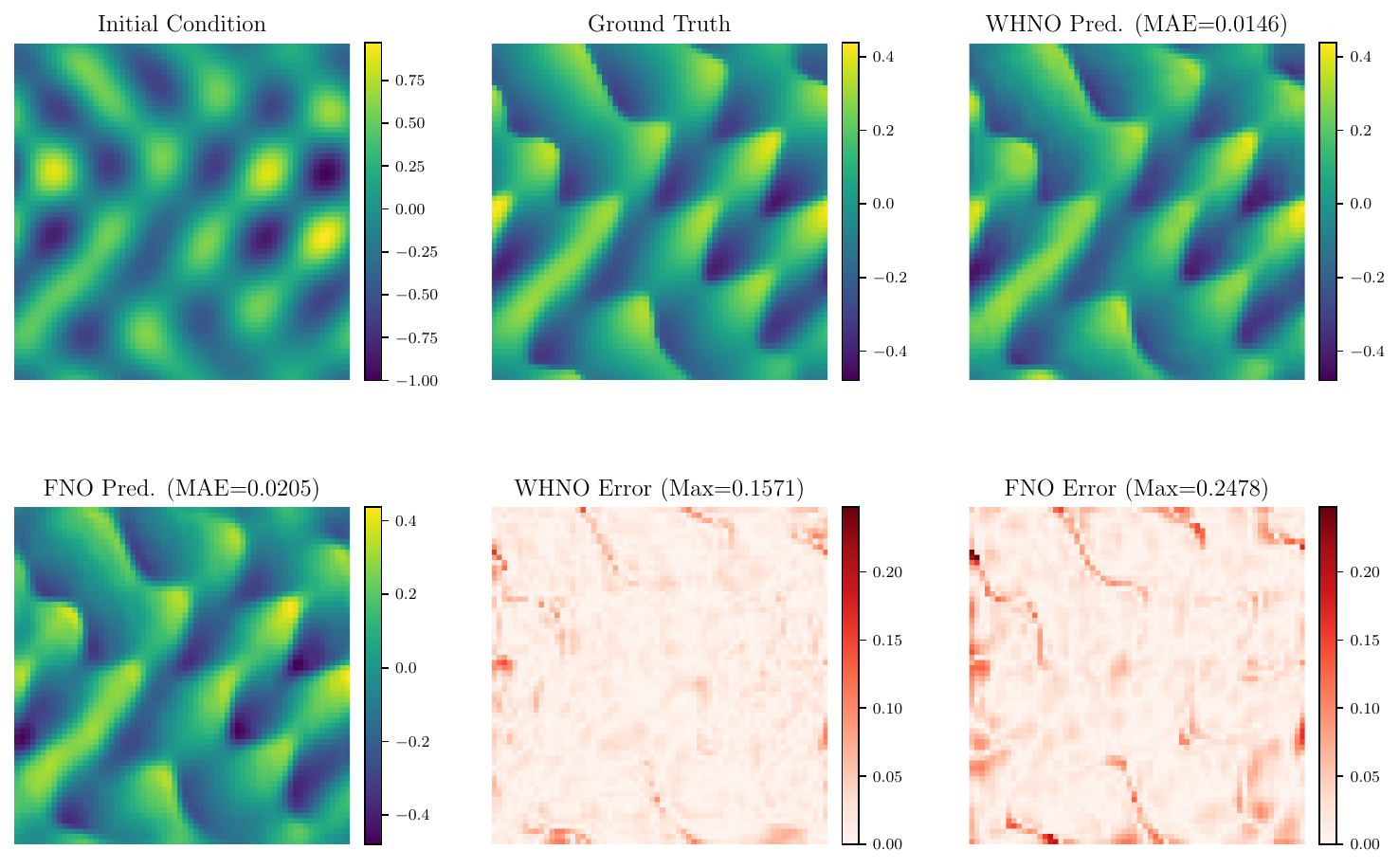}
\caption{2D Burgers equation with smooth sinusoidal initial conditions: WHNO and FNO predictions and absolute-error maps. The smooth initial condition (left) develops shock-like structure during evolution (centre); WHNO has lower error at those shocks.}
\label{fig:burgers_smooth}
\end{figure}

\begin{figure}[!htbp]
\centering
\includegraphics[width=\textwidth]{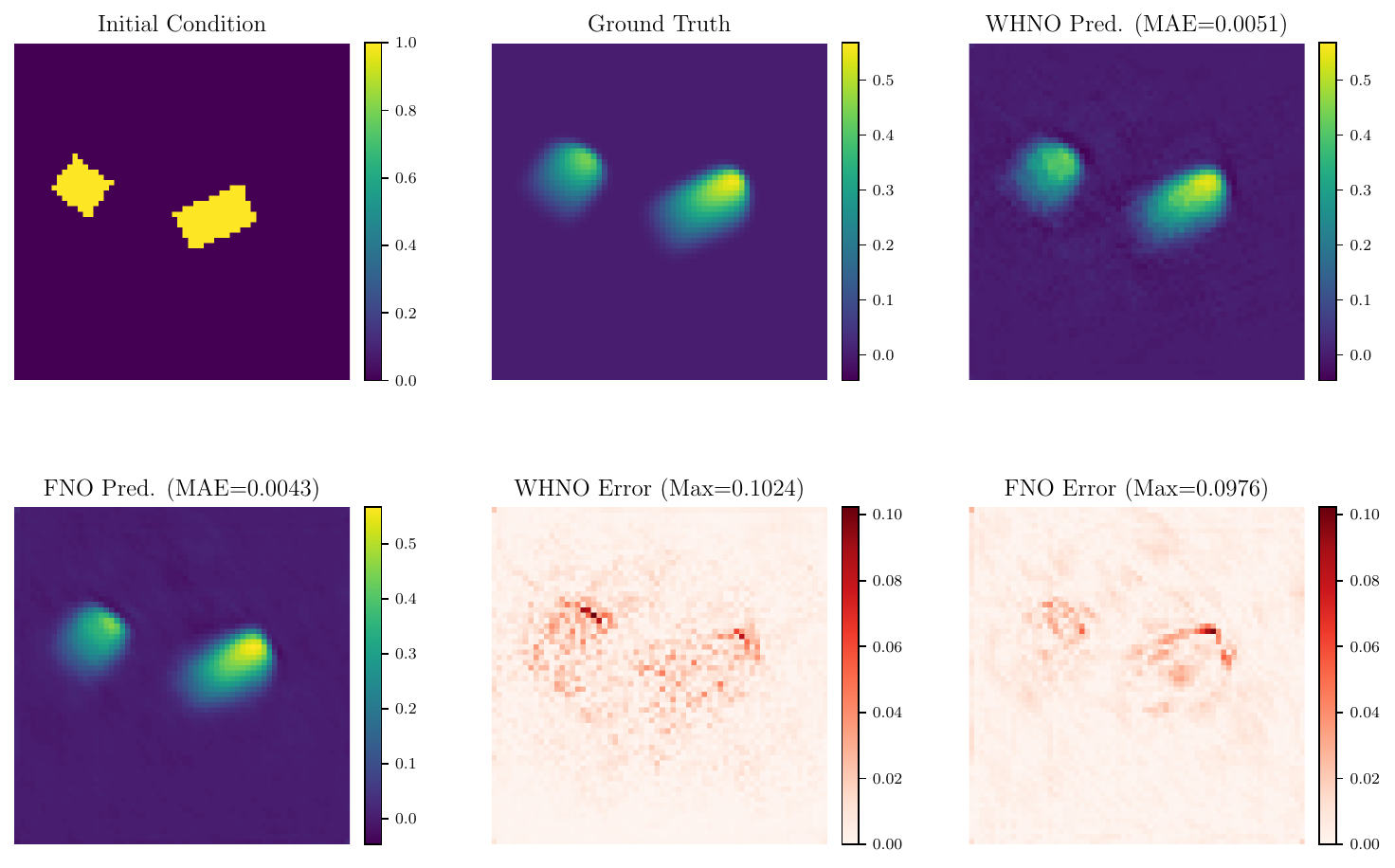}
\caption{2D Burgers equation with oblique-front initial conditions (interacting shocks at non-cardinal angles): WHNO and FNO predictions and absolute-error maps.}
\label{fig:burgers_oblique}
\end{figure}

Figures~\ref{fig:heat_offaxis_rotated}--\ref{fig:burgers_oblique} show prediction and error comparisons on representative samples from each family.

\subsection{Ensemble performance across all configurations}
\label{sec:ensemble_universal}

The cross-validated WHNO\,+\,FNO ensemble was fitted on each of the seven configurations reported in Sections~\ref{sec:results} and~\ref{sec:robustness}, using the same five-fold cross-validation protocol described earlier. Table~\ref{tab:universal_ensemble} reports the cross-validated $w^*$ and the per-fold mean MSE and $H^1$ at $w^*$, alongside the WHNO-only and FNO-only baselines evaluated on the same test folds.

\begin{table}[!htbp]
\centering
\small
\resizebox{\textwidth}{!}{%
\begin{tabular}{lc ccc ccc}
\toprule
& & \multicolumn{3}{c}{Test MSE} & \multicolumn{3}{c}{$H^1$ (gradient-MSE)} \\
\cmidrule(lr){3-5} \cmidrule(lr){6-8}
Configuration & $w^*$ & WHNO only & FNO only & Ensemble & WHNO only & FNO only & Ensemble \\
\midrule
Heat axis-aligned    & $0.572 \pm 0.016$ & $4.71 \times 10^{-4}$  & $5.66 \times 10^{-4}$  & $\mathbf{3.65 \times 10^{-4}}$ & $2.53 \times 10^{-4}$ & $2.64 \times 10^{-4}$ & $\mathbf{1.75 \times 10^{-4}}$ \\
Heat rotated         & $0.540 \pm 0.013$ & $4.96 \times 10^{-4}$  & $5.54 \times 10^{-4}$  & $\mathbf{3.53 \times 10^{-4}}$ & $2.87 \times 10^{-4}$ & $3.07 \times 10^{-4}$ & $\mathbf{1.89 \times 10^{-4}}$ \\
Heat circle          & $0.524 \pm 0.015$ & $6.52 \times 10^{-4}$  & $6.72 \times 10^{-4}$  & $\mathbf{4.76 \times 10^{-4}}$ & $3.66 \times 10^{-4}$ & $3.32 \times 10^{-4}$ & $\mathbf{2.36 \times 10^{-4}}$ \\
Heat Voronoi         & $0.632 \pm 0.037$ & $1.28 \times 10^{-3}$  & $1.59 \times 10^{-3}$  & $\mathbf{1.13 \times 10^{-3}}$ & $4.54 \times 10^{-4}$ & $4.34 \times 10^{-4}$ & $\mathbf{3.20 \times 10^{-4}}$ \\
Burgers block        & $0.648 \pm 0.020$ & $9.4 \times 10^{-5}$   & $1.59 \times 10^{-4}$  & $\mathbf{6.6 \times 10^{-5}}$  & $7.30 \times 10^{-5}$ & $9.22 \times 10^{-5}$ & $\mathbf{4.79 \times 10^{-5}}$ \\
Burgers smooth       & $0.744 \pm 0.008$ & $8.57 \times 10^{-4}$  & $1.44 \times 10^{-3}$  & $\mathbf{7.81 \times 10^{-4}}$ & $5.55 \times 10^{-4}$ & $8.58 \times 10^{-4}$ & $\mathbf{5.14 \times 10^{-4}}$ \\
Burgers oblique      & $0.520 \pm 0.013$ & $1.66 \times 10^{-4}$  & $1.78 \times 10^{-4}$  & $\mathbf{1.03 \times 10^{-4}}$ & $1.53 \times 10^{-4}$ & $1.11 \times 10^{-4}$ & $\mathbf{8.42 \times 10^{-5}}$ \\
\bottomrule
\end{tabular}%
}
\caption{Cross-validated WHNO\,+\,FNO ensemble performance across all seven (problem, geometry/IC) configurations evaluated in this paper. For each configuration $w^*$ is fitted on a held-out validation fold and metrics are computed on an independent test fold, repeated over five folds; the reported numbers are means over the five folds (standard deviations omitted for compactness; full per-fold detail is given in the supplementary CV outputs). In bold are the ensemble values, which are lower than both the WHNO-only and FNO-only entries in every row for both MSE and $H^1$.}
\label{tab:universal_ensemble}
\end{table}

The ensemble has strictly lower test MSE \emph{and} strictly lower $H^1$ than both single-model baselines in every one of the seven configurations. This includes the configurations where WHNO alone does not unambiguously beat FNO at the gradient level: on heat with disk inclusions and on Burgers with oblique fronts, FNO has lower single-model $H^1$ than WHNO, yet the ensemble at the cross-validated $w^*$ has lower $H^1$ than either model alone. The Walsh-Hadamard and Fourier representations cover partially complementary features of the reconstructed field, and combining them through a single cross-validated scalar weight gives an improvement on every problem variant we tested.

\section{Discussion}
\label{sec:discussion}

Walsh-Hadamard and Fourier representations are complementary rather than competitive. WHNO has lower MAE and lower $H^1$ than FNO at matched parameter count on both problems; weighted ensembles of WHNO and FNO have lower MSE still on both problems.

This complementarity follows from the spectral basis properties. The Walsh-Hadamard basis handles discontinuities directly because its rectangular wave functions are piecewise constant: they represent step discontinuities exactly without Gibbs ringing, allow aggressive spectral compression for piecewise constant fields, and align with multi-valued coefficient structures. The Fourier basis is more efficient on smooth variations, because sinusoidal functions capture gradual spatial transitions in few modes and have well-established convergence properties for differentiable functions.

In the PDE solutions considered here, both kinds of structure coexist. Heat conduction has sharp interfaces in the conductivity field but a smooth temperature field between them; the Burgers solutions have shocks but also large mostly-smooth regions. Weighted ensembles use both bases, and the cross-validated weight indicates how the two contribute on a given problem: $w^* = 0.572$ for heat conduction, $w^* = 0.648$ for the Burgers equation.

How much an ensemble improves over the better of the two single models depends on how informative the second basis is relative to the first. There is an empirical pattern across the seven configurations in Table~\ref{tab:universal_ensemble}: configurations in which the WHNO-only and FNO-only test MSEs are similar tend to gain more from ensembling than configurations in which one of the two pure methods clearly dominates. Quantitatively, the four configurations whose single-model MSE ratio $\min(\mathrm{WHNO},\mathrm{FNO})/\max(\mathrm{WHNO},\mathrm{FNO})$ lies in $[0.83, 0.97]$ (i.e.\ the two single models are within roughly $20\%$ of each other) have a mean ensemble improvement over the better single model of $29\%$, while the three configurations whose ratio lies in $[0.59, 0.80]$ have a mean improvement of $17\%$. The rank correlation between similarity and ensemble gain across the seven configurations is positive but moderate (Spearman $\rho \approx 0.32$), and there is at least one clear counter-example: on the Burgers block-IC case WHNO has a $\sim 70\%$ lower MSE than FNO, yet the ensemble still reduces MSE by a further $30\%$ over WHNO alone. The pattern is therefore a trend, not a strict rule.

A useful intuition is that the two bases recover partially complementary features of the reconstructed field. When the two single models have comparable accuracy on a given configuration the second basis tends to contribute non-trivially in regions where the first basis is weak, so a near-equal blend ($w^* \approx 0.5$) closes a larger fraction of the residual error. When one single model is already much better than the other, the second basis can still help, but the room left to gain is smaller and $w^*$ shifts accordingly. We do not have a closed-form prediction for $w^*$ in advance of the cross-validation; the present data is consistent with $w^*$ shifting in the direction of the better single model, but a single-seed campaign on two problems is not enough to make this prediction quantitative.

\begin{figure}[!htbp]
\centering
\includegraphics[width=\textwidth]{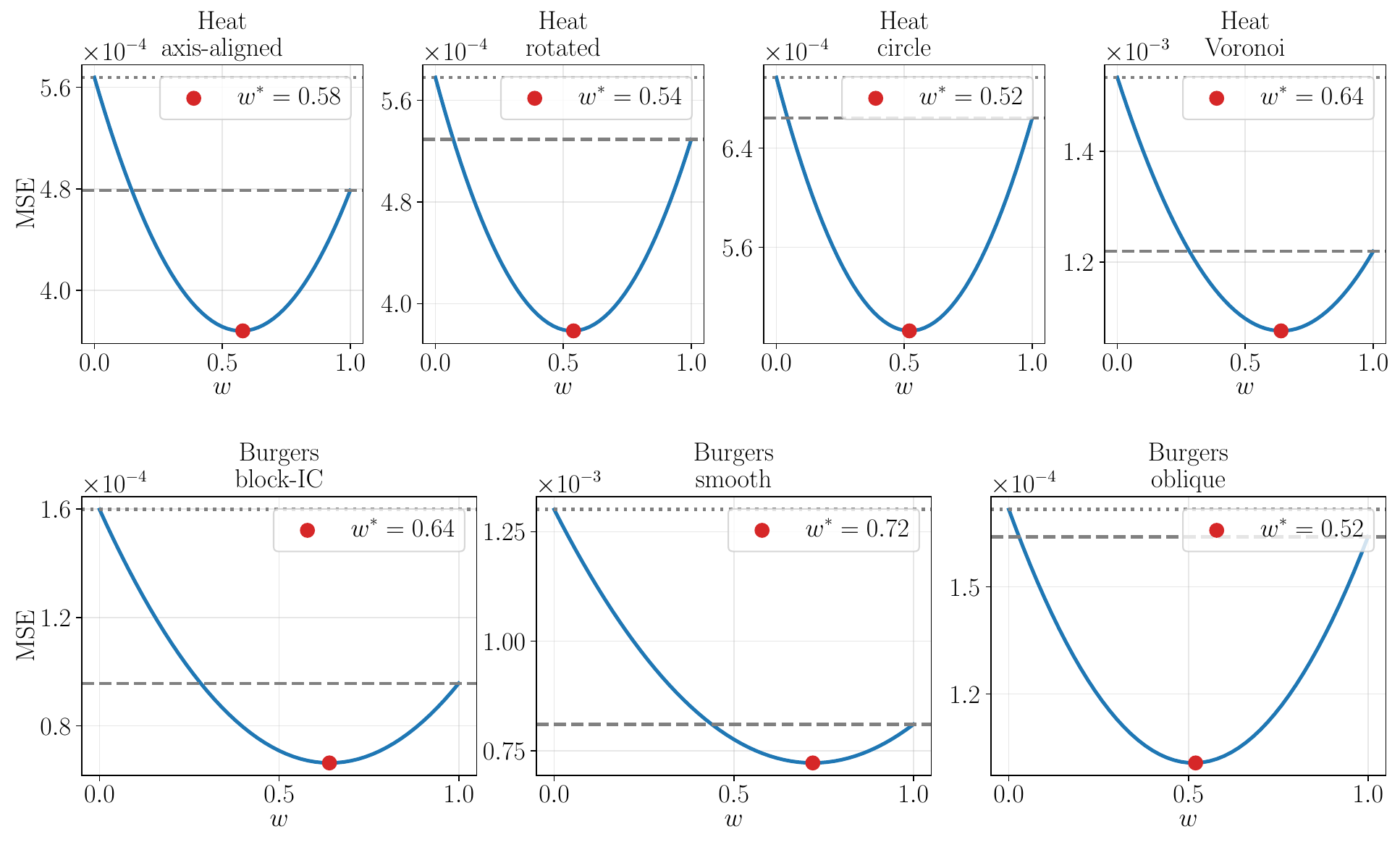}
\caption{Validation MSE as a function of WHNO weight $w$ for each of the seven (problem, geometry/IC) pairs. Dashed line: WHNO-only baseline; dotted line: FNO-only baseline; red marker: argmin $w^*$. The location of $w^*$ tracks the relative single-model performance.}
\label{fig:weight_optimization_all}
\end{figure}

Figure~\ref{fig:weight_optimization_all} shows the MSE-versus-$w$ sweep on a validation set for each of the seven (problem, geometry/IC) pairs reported in this paper. The location of $w^*$ moves with the relative single-model performance: pairs in which WHNO and FNO are close on the validation MSE (heat axis-aligned, rotated, circle, and Burgers oblique) have $w^*$ near $0.5$; pairs in which WHNO has a clear single-model edge (Burgers smooth, Burgers block-IC, heat Voronoi) have $w^*$ shifted toward WHNO.

WHNO, FNO, and their ensembles all have $O(n \log n)$ computational complexity. The Fast Walsh-Hadamard Transform matches FFT efficiency, so WHNO has the same per-pass cost as FNO. Ensembles require two forward passes (WHNO+FNO) at inference time, doubling inference cost relative to a single model but requiring no additional training. Practical deployment can cache both model predictions for multiple downstream analyses.

Our findings on heat conduction and the Burgers equation suggest when to use each approach. WHNO alone is appropriate when the coefficients are predominantly discontinuous (binary permeability, sharp conductivity contrasts), when only a single forward pass can be afforded at inference, or when training two models is not practical. The WHNO+FNO ensemble is appropriate when the application can tolerate the doubled inference cost, when both discontinuous and smooth structure are present, and when a held-out validation set is available for fitting $w^*$. FNO alone is appropriate when the coefficients are smooth and continuous, when problems lack sharp interfaces, or when an FNO implementation is already deployed.

The ensemble weight $w$ is tuned via cross-validation: $w^*$ is fitted on a held-out validation fold and the metrics reported in Tables~\ref{tab:ensemble_heat} and~\ref{tab:ensemble_burgers} are computed on an independent test fold, with the procedure repeated over five folds. We find $w^* = 0.572 \pm 0.016$ for heat conduction and $w^* = 0.648 \pm 0.020$ for the Burgers equation.

Our work supports two principles for neural operator design. First, the spectral basis matters at matched parameter count: replacing Fourier with Walsh-Hadamard reduces MAE and $H^1$ on the two baseline problems considered here, at the same per-pass cost. Second, and we view this as the central practical contribution of the paper: combining the two bases through a cross-validated scalar weight gives an improvement over both WHNO alone and FNO alone on \emph{every} configuration we evaluated (four heat geometries and three Burgers initial-condition families; Table~\ref{tab:universal_ensemble}). This includes the configurations where WHNO alone does not have a clear single-model edge over FNO at the gradient level, such as heat with disk inclusions or Burgers with oblique fronts. The Walsh-Hadamard and Fourier bases recover partially complementary features of the reconstructed field, and ensembling them produces a method that improves on a Fourier-only baseline irrespective of which single basis happens to dominate on a given configuration.

\subsection{Limitations and Scope}
\label{sec:limitations}

We acknowledge the following limitations, which also delimit the regime in which WHNO is the right tool.

\emph{Dyadic grid restriction.} The Fast Walsh-Hadamard Transform requires both spatial dimensions to be powers of two; consequently WHNO is structurally tied to regular Cartesian grids at power-of-two resolution. Extension to non-dyadic grids requires either interpolation (which loses geometric fidelity at material boundaries) or a fundamentally different transform, such as graph wavelets. This is a deliberate design choice that reflects the trade-off implicit in choosing a structured spectral basis; for fully unstructured or irregular geometries we recommend the graph-based neural operators of \cite{li2020neural,li2020multipole} as more natural alternatives.

\emph{Sensitivity to basis--geometry alignment.} The discontinuities in our baseline experiments (axis-aligned rectangular inclusions in heat conduction and axis-aligned blocks in Burgers) align with the Walsh-Hadamard basis cell boundaries, which is a configuration favourable to WHNO. The robustness experiments in Section~\ref{sec:robustness} show that the WHNO advantage on $L^p$ metrics carries over to rotated rectangles, disk inclusions, Voronoi tessellations, smooth sinusoidal initial conditions, and oblique-front initial conditions, but with reduced magnitude on the non-aligned cases; on the $H^1$ metric the comparison is mixed, with FNO ahead of WHNO on disks, on Voronoi, and on oblique-front Burgers. The Walsh-Hadamard advantage is therefore real but configuration-dependent: it is largest when the discontinuity structure can be expressed compactly in the basis, and shrinks (or reverses, at the gradient level) when it cannot.

\emph{Comparison scope.} Our controlled comparison is specifically between WHNO and FNO with all non-spectral architectural components held fixed, which isolates the contribution of the spectral basis itself. Comparison against other operator-learning architectures with different decoders or attention mechanisms---U-shaped Neural Operators (UNO)~\cite{rahman2022u}, Galerkin Transformer~\cite{cao2021choosing}, hierarchical or multi-wavelet operators~\cite{gupta2021neurips}---would conflate spectral-basis effects with decoder-architecture effects, and we leave such broader benchmarking to future work. Physics-informed neural networks~\cite{raissi2019physics} address a different problem regime (per-instance optimisation rather than operator learning) and are not directly comparable as drop-in surrogates.

The Wavelet Neural Operator (WNO)~\cite{tripura2022wavelet} is closer to WHNO both ideologically and structurally: both replace the Fourier transform with a basis built from step-like or locally supported functions, and both operate on truncated spectral coefficients. The wavelet construction, however, introduces several modelling choices that the Walsh-Hadamard transform does not have:

\begin{itemize}[leftmargin=*,itemsep=2pt]
  \item \emph{Wavelet family} --- Haar, Daubechies-$N$, Symlets, Coiflets, biorthogonal pairs, with qualitatively different localisation, smoothness, and vanishing-moment properties.
  \item \emph{Number of decomposition levels} --- how deep the wavelet tree goes, controlling which scales are explicitly represented versus collapsed into the coarsest-level coefficients.
  \item \emph{Boundary handling} --- periodic, symmetric, reflective, or zero-padded extension, with non-trivial effects near grid boundaries.
  \item \emph{Filter length} --- longer filters give more vanishing moments at the cost of locality.
\end{itemize}

The Walsh-Hadamard transform has none of these knobs: a single canonical basis (modulo coefficient ordering, which leaves the truncation rule invariant) with one cutoff parameter $k$, structurally identical to FNO's mode count. A WNO implementation must commit to a family-and-levels combination before training, and the appropriate choice depends on assumptions about the regularity of the target operator that may not be known a priori.

To put the closest WNO analog side by side with WHNO at matched conditions we ran a controlled experiment with single-level Haar wavelets --- the multi-scale cousin of the Walsh basis, since both are built from rectangular step functions --- with $k=32$ chosen so the spectral compression matches WHNO's sequency truncation exactly, identical encoder/decoder structure, and identical parameter count ($1{,}555{,}153$). On the four heat-conduction geometries the Haar-WNO outperforms WHNO at the single-model level ($-15\%$ to $-27\%$ MAE) and the FNO+Haar-WNO cross-validated ensemble beats the FNO+WHNO ensemble of Table~\ref{tab:universal_ensemble} ($-30\%$ to $-43\%$ MSE). On the three Burgers configurations the picture inverts: Haar-WNO is markedly worse than WHNO at the single-model level ($+27\%$ to $+67\%$ MAE) and the FNO+Haar-WNO ensemble loses to FNO+WHNO ($+34\%$ to $+87\%$ MSE). The split has a clean structural explanation: the heat solution is determined locally by the conductivity neighbourhood, which fits the local support of a single-level Haar basis well; the Burgers solution at $T = 0.25$ depends on initial-condition information from a region whose radius grows with the $500$ advection time steps, and the global support of the Walsh basis encodes that long-range dependence in low-sequency coefficients that a single-scale local wavelet has no mechanism to express. A multi-level Haar tree (or a Daubechies-$N$ family with longer filter support) would likely close part of the Burgers gap, but tuning the family-and-levels pair to recover that depends on prior knowledge of the target operator's regularity --- the same modelling degree of freedom that WHNO removes. Full numerical detail is in Appendix~\ref{app:wno}.

\section{Conclusion}
\label{sec:conclusion}

We introduced the Walsh-Hadamard Neural Operator (WHNO), a neural operator architecture that uses Walsh-Hadamard transforms. These rectangular wave basis functions are well-suited for discontinuous coefficients and are combined with learnable spectral weights that capture global dependencies. Beyond showing WHNO's effectiveness, our key finding is that weighted ensembles of WHNO and FNO achieve improvements over either model alone, which reveals a systematic complementarity between their spectral representations.

On heat conduction with discontinuous conductivity, WHNO at matched parameter count has lower MAE than FNO ($0.0153$ vs.\ $0.0166$ over $100$ test samples) and lower $H^1$ error. On the 2D Burgers equation with discontinuous initial conditions, WHNO again has lower MAE than FNO ($0.0064$ vs.\ $0.0084$) and lower $H^1$ error. In both cases the FNO errors are concentrated more sharply at the discontinuities, consistent with Gibbs-type behaviour, while the WHNO errors are spread more evenly across the domain.

Weighted ensembles of WHNO and FNO improve upon both single models on both problems. The ensemble framework is straightforward: train WHNO and FNO independently, then fit $w^* \in [0,1]$ on held-out validation data so as to minimise $\mathbb{E}[(w \cdot u_{\text{WHNO}} + (1-w) \cdot u_{\text{FNO}} - u_{\text{true}})^2]$. Under five-fold cross-validation we find $w^* = 0.572 \pm 0.016$ for heat conduction and $w^* = 0.648 \pm 0.020$ for the Burgers equation; both values keep a non-trivial FNO contribution. Ensembling doubles inference cost but requires no additional training.

Our work supports two principles. First, the spectral basis matters at matched parameter count: replacing Fourier with Walsh-Hadamard reduces MAE and $H^1$ on the two baseline problems, at the same per-pass cost. Second, and we view this as the most useful practical message of the paper: combining the two bases through a single cross-validated scalar weight gives strictly lower test MSE and strictly lower $H^1$ than \emph{both} WHNO alone and FNO alone on every one of the seven (problem, geometry/IC) configurations we evaluated, including the configurations where WHNO alone does not unambiguously beat FNO. The cross-validated weight tracks the relative single-model performance of the two bases on a given configuration, and the ensemble improvement is consistent across all seven cases. A practitioner who already has a Fourier-only baseline therefore has, in WHNO\,+\,FNO ensembling, a low-risk method to obtain a measurable accuracy improvement on PDE problems with discontinuous structure.

Practitioners should use WHNO alone when computational budget is tight and strong discontinuities are present. The WHNO+FNO ensemble is appropriate when the application can afford a doubled inference cost and a held-out validation set is available for fitting $w^*$.

Future work could explore several directions: extension to three-dimensional geometries and realistic microstructures; adaptive ensemble weighting (learning $w(\mathbf{x})$ spatially rather than as a scalar); a theoretical analysis of approximation rates for Walsh-based operators and of ensemble generalisation bounds; and ways to reduce inference time, for example by sharing intermediate features between the two models.

In conclusion, the Walsh-Hadamard Neural Operator and its ensemble with FNO demonstrate that the choice of spectral basis is a useful design axis for neural operators on PDEs with discontinuous coefficients or initial conditions, and that cross-validated weighted combinations of complementary bases can further improve accuracy at predictable additional cost.

\section*{Declaration of Competing Interest}
The authors declare that they have no known competing financial interests or personal relationships that could have appeared to influence the work reported in this paper.

\section*{Data Availability}
All code, trained models, and data needed to reproduce the results presented in this paper are publicly available at \url{https://github.com/gmcavallazzi/WHNO}. The repository includes the implementation of the WHNO, training scripts for all experiments (Darcy flow, heat conduction, Burgers equation), pre-trained model weights, data generation scripts, and instructions for reproducing all figures and results.

\appendix

\section{Wavelet Neural Operator comparison}
\label{app:wno}

This appendix records the controlled WHNO vs.\ Haar-WNO comparison summarised in the \emph{Comparison scope} subsection of Section~\ref{sec:limitations}.

\subsection*{Haar-WNO architecture}

The Haar-WNO model used here is the closest direct analog of WHNO: the only architectural change relative to WHNO is the spectral transform inside the encoder. Specifically, the encoder applies a single-level 2D orthonormal Haar wavelet decomposition to the input, keeps the LL (low-pass) band, multiplies it by a learnable spectral weight tensor of shape $k \times k \times c_{\text{in}} \times c_{\text{out}}$ identical to the WHNO encoder, zeros the LH, HL, and HH detail bands at the output channel count, and applies the inverse Haar DWT. With input height $H = 64$ and $k = H/2 = 32$, the LL band size matches WHNO's sequency truncation exactly. All non-spectral components (dilated-convolution decoder, residual structure, input-field re-injection, GELU activations, batch normalisation, training protocol) are identical to the WHNO models reported in Section~\ref{sec:results}. The total parameter count is $1{,}555{,}153$ at the configuration used for the heat and Burgers experiments, identical to WHNO.

\subsection*{Single-model three-way comparison}

Table~\ref{tab:wno_singlemodel} reports mean absolute, maximum, mean-squared, and $H^1$ errors for WHNO, FNO, and Haar-WNO on $100$ independent test samples (seed $999$) for each of the seven (problem, geometry/IC) configurations of Section~\ref{sec:results}.

\begin{table}[!htbp]
\centering
\small
\begin{tabular}{l l c c c c}
\toprule
Configuration & Model & MAE & Max & MSE & $H^1$ \\
\midrule
\multirow{3}{*}{Heat axis-aligned}
  & WHNO     & $0.01525$         & $0.154$         & $4.63 \times 10^{-4}$         & $2.50 \times 10^{-4}$ \\
  & FNO      & $0.01656$         & $0.165$         & $5.57 \times 10^{-4}$         & $2.62 \times 10^{-4}$ \\
  & Haar-WNO & $\mathbf{0.01290}$ & $0.170$         & $\mathbf{3.48 \times 10^{-4}}$ & $\mathbf{2.31 \times 10^{-4}}$ \\
\midrule
\multirow{3}{*}{Heat rotated}
  & WHNO     & $0.01543$         & $0.235$         & $5.20 \times 10^{-4}$         & $2.94 \times 10^{-4}$ \\
  & FNO      & $0.01671$         & $0.192$         & $5.83 \times 10^{-4}$         & $3.11 \times 10^{-4}$ \\
  & Haar-WNO & $\mathbf{0.01120}$ & $\mathbf{0.177}$ & $\mathbf{3.11 \times 10^{-4}}$ & $\mathbf{2.02 \times 10^{-4}}$ \\
\midrule
\multirow{3}{*}{Heat circle}
  & WHNO     & $0.01823$         & $0.208$         & $6.79 \times 10^{-4}$         & $3.71 \times 10^{-4}$ \\
  & FNO      & $0.01864$         & $0.196$         & $7.11 \times 10^{-4}$         & $3.39 \times 10^{-4}$ \\
  & Haar-WNO & $\mathbf{0.01460}$ & $\mathbf{0.176}$ & $\mathbf{4.70 \times 10^{-4}}$ & $\mathbf{2.68 \times 10^{-4}}$ \\
\midrule
\multirow{3}{*}{Heat Voronoi}
  & WHNO     & $0.02575$         & $0.295$         & $1.30 \times 10^{-3}$         & $4.57 \times 10^{-4}$ \\
  & FNO      & $0.02781$         & $0.305$         & $1.56 \times 10^{-3}$         & $4.35 \times 10^{-4}$ \\
  & Haar-WNO & $\mathbf{0.01874}$ & $0.321$         & $\mathbf{7.00 \times 10^{-4}}$ & $\mathbf{3.90 \times 10^{-4}}$ \\
\midrule
\multirow{3}{*}{Burgers block-IC}
  & WHNO     & $\mathbf{0.00661}$ & $\mathbf{0.190}$ & $\mathbf{1.05 \times 10^{-4}}$ & $\mathbf{8.28 \times 10^{-5}}$ \\
  & FNO      & $0.00878$         & $0.183$         & $1.86 \times 10^{-4}$         & $1.07 \times 10^{-4}$ \\
  & Haar-WNO & $0.01103$         & $0.315$         & $3.47 \times 10^{-4}$         & $2.79 \times 10^{-4}$ \\
\midrule
\multirow{3}{*}{Burgers smooth}
  & WHNO     & $\mathbf{0.01909}$ & $\mathbf{0.648}$ & $\mathbf{1.01 \times 10^{-3}}$ & $\mathbf{6.07 \times 10^{-4}}$ \\
  & FNO      & $0.02533$         & $0.651$         & $1.61 \times 10^{-3}$         & $9.08 \times 10^{-4}$ \\
  & Haar-WNO & $0.02863$         & $0.693$         & $2.02 \times 10^{-3}$         & $1.27 \times 10^{-3}$ \\
\midrule
\multirow{3}{*}{Burgers oblique}
  & WHNO     & $\mathbf{0.00812}$ & $0.272$         & $\mathbf{1.67 \times 10^{-4}}$ & $1.54 \times 10^{-4}$ \\
  & FNO      & $0.00845$         & $\mathbf{0.245}$ & $1.77 \times 10^{-4}$         & $\mathbf{1.10 \times 10^{-4}}$ \\
  & Haar-WNO & $0.01035$         & $0.347$         & $3.02 \times 10^{-4}$         & $2.66 \times 10^{-4}$ \\
\bottomrule
\end{tabular}
\caption{Single-model three-way comparison at matched parameter count ($1{,}555{,}153$ per model) on $100$ independent test samples (seed $999$). Bold entries mark the lowest value in each metric column within each configuration. Haar-WNO wins on every heat configuration; WHNO wins on every Burgers configuration.}
\label{tab:wno_singlemodel}
\end{table}

\subsection*{Cross-validated FNO ensemble comparison}

Table~\ref{tab:wno_ensemble} reports the cross-validated FNO+Haar-WNO ensemble alongside the FNO+WHNO ensemble from Table~\ref{tab:universal_ensemble} of the main text, using the same five-fold protocol.

\begin{table}[!htbp]
\centering
\small
\begin{tabular}{l c c c c c}
\toprule
& \multicolumn{2}{c}{FNO + Haar-WNO} & \multicolumn{2}{c}{FNO + WHNO (Table~\ref{tab:universal_ensemble})} & $\Delta$ MSE \\
\cmidrule(lr){2-3} \cmidrule(lr){4-5}
              &                     & test MSE              &                     & test MSE              & (Haar vs.\\
Configuration & $w^*_{\text{Haar}}$ & ($\times 10^{-4}$)    & $w^*_{\text{WHNO}}$ & ($\times 10^{-4}$)    & WHNO) \\
\midrule
Heat axis-aligned & $0.644 \pm 0.015$ & $2.57$  & $0.572 \pm 0.016$ & $3.65$  & $\mathbf{-30\%}$ \\
Heat rotated      & $0.672 \pm 0.010$ & $2.29$  & $0.540 \pm 0.013$ & $3.53$  & $\mathbf{-35\%}$ \\
Heat circle       & $0.624 \pm 0.008$ & $3.13$  & $0.524 \pm 0.015$ & $4.76$  & $\mathbf{-34\%}$ \\
Heat Voronoi      & $0.800 \pm 0.013$ & $6.41$  & $0.632 \pm 0.037$ & $11.32$ & $\mathbf{-43\%}$ \\
Burgers block-IC  & $0.320 \pm 0.013$ & $1.24$  & $0.648 \pm 0.020$ & $0.66$  & $+87\%$ \\
Burgers smooth    & $0.348 \pm 0.020$ & $13.44$ & $0.744 \pm 0.008$ & $7.81$  & $+72\%$ \\
Burgers oblique   & $0.344 \pm 0.008$ & $1.38$  & $0.520 \pm 0.013$ & $1.03$  & $+34\%$ \\
\bottomrule
\end{tabular}
\caption{Five-fold cross-validated FNO+Haar-WNO ensemble versus the FNO+WHNO ensemble of Table~\ref{tab:universal_ensemble} at matched parameter count. All test-MSE values are multiplied by $10^{4}$. The optimum weight $w^*$ shifts toward the better single model on each configuration: toward Haar on heat (weights $0.62$--$0.80$), toward FNO on Burgers (weights $0.32$--$0.35$). The $\Delta$ MSE column reports the relative change of the FNO+Haar ensemble against the FNO+WHNO ensemble.}
\label{tab:wno_ensemble}
\end{table}

\subsection*{Interpretation}

Across the seven configurations the result splits strictly along the static-versus-dynamic axis. The Haar wavelet basis wins on every heat-conduction geometry, where the quasi-steady temperature is determined locally by the conductivity neighbourhood; the Walsh-Hadamard basis wins on every Burgers configuration, where the $500$-step advective evolution couples distant regions of the domain and benefits from globally supported basis functions. The single-level Haar transform was deliberately chosen to give the closest WHNO analog (matched coefficient grid, matched parameter count, matched encoder/decoder structure), but it represents a particular point in the wavelet design space; a multi-level Haar decomposition or a Daubechies family with longer filters might shift the static-vs.-dynamic balance, at the cost of adding hyperparameters that WHNO does not have. The takeaway is consistent with the discussion in Section~\ref{sec:limitations}: spectral-basis choice matters, the right wavelet construction depends on the structure of the target operator, and WHNO is the basis-choice-free option whose performance is competitive across both regimes either alone or in the FNO ensemble.


\begin{thebibliography}{10}
\expandafter\ifx\csname url\endcsname\relax
  \def\url#1{\texttt{#1}}\fi
\expandafter\ifx\csname urlprefix\endcsname\relax\def\urlprefix{URL }\fi
\expandafter\ifx\csname href\endcsname\relax
  \def\href#1#2{#2} \def\path#1{#1}\fi

\bibitem{kovachki2021neural}
N.~Kovachki, Z.~Li, B.~Liu, K.~Azizzadenesheli, K.~Bhattacharya, A.~Stuart,
  A.~Anandkumar, Neural operator: Learning maps between function spaces with
  applications to pdes, Journal of Machine Learning Research 24~(89) (2023)
  1--97.

\bibitem{li2020fourier}
Z.~Li, N.~Kovachki, K.~Azizzadenesheli, B.~Liu, K.~Bhattacharya, A.~Stuart,
  A.~Anandkumar, Fourier neural operator for parametric partial differential
  equations, arXiv preprint arXiv:2010.08895 (2020).

\bibitem{li2021fourier}
Z.~Li, D.~Z. Huang, B.~Liu, A.~Anandkumar, {Fourier neural operator with
  learned deformations for pdes on general geometries}, {Journal of Machine
  Learning Research} 24~(388) (2023) 1--26.

\bibitem{ochoa-tapia1995momentum}
J.~Ochoa-Tapia, S.~Whitaker, Momentum transfer at the boundary between a porous
  medium and a homogeneous fluid—ii. comparison with experiment,
  International Journal of Heat and Mass Transfer 38~(14) (1995) 2647--2655.

\bibitem{tien1994challenges}
C.~L. Tien, G.~Chen, Challenges in microscale conductive and radiative heat
  transfer, ASME. J. Heat Transfer 116~(4) (1994) 799--807.
\newblock \href {https://doi.org/10.1115/1.2911450}
  {\path{doi:10.1115/1.2911450}}.

\bibitem{kwon2021review}
Y.~Kwon, J.~Park, Y.~Jeon, J.~Hong, H.~Park, J.~Lee, A review of polymer
  composites based on carbon fillers for thermal management applications:
  Design, preparation, and properties, Polymers 13~(8) (2021) 1312.

\bibitem{skews1967perturbed}
B.~Skews, The perturbed region behind a diffracting shock wave, Journal of
  Fluid Mechanics 29~(4) (1967) 705--719.

\bibitem{canuto2006spectral}
C.~Canuto, M.~Hussaini, A.~Quarteroni, T.~Zang, Spectral methods: fundamentals
  in single domains, Springer Berlin Heidelberg, Berlin, Heidelberg, 2006.

\bibitem{fine1949walsh}
N.~J. Fine, On the walsh functions, Transactions of the American Mathematical
  Society 65~(3) (1949) 372--414.

\bibitem{lu2021learning}
L.~Lu, P.~Jin, G.~Pang, Z.~Zhang, G.~E. Karniadakis, Learning nonlinear
  operators via deeponet based on the universal approximation theorem of
  operators, Nature Machine Intelligence 3~(3) (2021) 218--229.

\bibitem{tran2021factorized}
A.~Tran, A.~Mathews, L.~Xie, C.~S. Ong, Factorized fourier neural operators,
  arXiv preprint arXiv:2111.13802 (2021).

\bibitem{cao2021choosing}
S.~Cao, Choose a transformer: Fourier or galerkin, Advances in Neural
  Information Processing Systems 34 (2021) 24924--24940.

\bibitem{li2022fourier}
Z.~Li, D.~Z. Huang, B.~Liu, A.~Anandkumar, Fourier neural operator with learned
  deformations for pdes on general geometries, Journal of Computational Physics
  471 (2022) 111617.

\bibitem{pathak2022fourcastnet}
J.~Pathak, S.~Subramanian, P.~Harrington, S.~Raja, A.~Chattopadhyay,
  M.~Mardani, et~al., Fourcastnet: A global data-driven high-resolution weather
  model using adaptive fourier neural operators, arXiv preprint
  arXiv:2202.11214 (2022).

\bibitem{carslaw1906theory}
{Carslaw, Horatio S}, {The Theory of Fourier’s Series and Integrals},
  {Nature} 75~(1931) (1906) 14--14.

\bibitem{raissi2019physics}
M.~Raissi, P.~Perdikaris, G.~E. Karniadakis, Physics-informed neural networks:
  A deep learning framework for solving forward and inverse problems involving
  nonlinear partial differential equations, Journal of Computational Physics
  378 (2019) 686--707.

\bibitem{karniadakis2021physics}
G.~E. Karniadakis, I.~G. Kevrekidis, L.~Lu, P.~Perdikaris, S.~Wang, L.~Yang,
  Physics-informed machine learning, Nature Reviews Physics 3~(6) (2021)
  422--440.

\bibitem{mao2020physics}
Z.~Mao, A.~D. Jagtap, G.~E. Karniadakis, Physics-informed neural networks for
  high-speed flows, Computer Methods in Applied Mechanics and Engineering 360
  (2020) 112789.

\bibitem{gupta2021neurips}
G.~Gupta, X.~Xiao, P.~Bogdan, Multiwavelet-based operator learning for
  differential equations, in: Advances in Neural Information Processing Systems
  (NeurIPS), Vol.~34, 2021, pp. 24048--24062.

\bibitem{walsh1923closed}
J.~L. Walsh, A closed set of normal orthogonal functions, American Journal of
  Mathematics 45~(1) (1923) 5--24.

\bibitem{beauchamp1975walsh}
K.~G. Beauchamp, Walsh functions and their applications, Academic Press, 1975.

\bibitem{fino1976unified}
B.~J. Fino, V.~R. Algazi, Unified matrix treatment of the fast walsh-hadamard
  transform, IEEE Transactions on Computers 100~(11) (1976) 1142--1146.

\bibitem{shanmugam1979walsh}
K.~S. Shanmugam, A.~M. Breipohl, Walsh-hadamard transform for image coding,
  Proceedings of the IEEE 67~(7) (1979) 1025--1026.

\bibitem{li2020neural}
Z.~Li, N.~Kovachki, K.~Azizzadenesheli, B.~Liu, K.~Bhattacharya, A.~Stuart,
  A.~Anandkumar, Neural operator: Graph kernel network for partial differential
  equations, arXiv preprint arXiv:2003.03485 (2020).

\bibitem{li2020multipole}
Z.~Li, N.~Kovachki, K.~Azizzadenesheli, B.~Liu, A.~Stuart, K.~Bhattacharya,
  A.~Anandkumar, Multipole graph neural operator for parametric partial
  differential equations, Advances in Neural Information Processing Systems 33
  (2020) 6755--6766.

\bibitem{mo2019deep}
S.~Mo, Y.~Zhu, N.~Zabaras, X.~Shi, J.~Wu, Deep convolutional encoder-decoder
  networks for uncertainty quantification of dynamic multiphase flow in
  heterogeneous media, Water Resources Research 55~(1) (2019) 703--728.

\bibitem{zhu2019physics}
Y.~Zhu, N.~Zabaras, Bayesian deep convolutional encoder-decoder networks for
  surrogate modeling and uncertainty quantification, Journal of Computational
  Physics 366 (2018) 415--447.

\bibitem{raissi2018numerical}
M.~Raissi, P.~Perdikaris, G.~E. Karniadakis, Numerical gaussian processes for
  time-dependent and nonlinear partial differential equations, SIAM Journal on
  Scientific Computing 40~(1) (2018) A172--A198.

\bibitem{cai2021physics}
S.~Cai, Z.~Mao, Z.~Wang, M.~Yin, G.~E. Karniadakis, Physics-informed neural
  networks (pinns) for heat transfer problems, Journal of Heat Transfer 143~(6)
  (2021) 060801.

\bibitem{kashinath2021physics}
K.~Kashinath, M.~Mustafa, A.~Albert, J.~Wu, C.~Jiang, S.~Esmaeilzadeh, et~al.,
  Physics-informed machine learning: Case studies for weather and climate
  modelling, Philosophical Transactions of the Royal Society A 379~(2194)
  (2021) 20200093.

\bibitem{hendrycks2016gaussian}
D.~Hendrycks, Gaussian error linear units (gelus), arXiv preprint
  arXiv:1606.08415 (2016).

\bibitem{rahman2022u}
M.~A. Rahman, Z.~E. Ross, K.~Azizzadenesheli, {U-NO}: {U}-shaped neural
  operators, Transactions on Machine Learning ResearchArXiv:2204.11127 (2023).

\bibitem{tripura2022wavelet}
T.~Tripura, S.~Chakraborty, Wavelet neural operator for solving parametric
  partial differential equations in computational mechanics problems, Computer
  Methods in Applied Mechanics and Engineering 404 (2023) 115783.

\end{thebibliography}
\end{document}